\documentclass{article}%
\pdfoutput=1
\newcommand{\sv}[1]{}%
 \newcommand{\lv}[1]{#1}%
\usepackage{etoolbox}

 \newcommand{\toappendix}[1]{#1}%
\usepackage[T1]{fontenc}
\usepackage[utf8]{inputenc}
\usepackage{graphicx} %
\usepackage{subcaption}
\usepackage{lmodern}
\lv{\usepackage{a4wide}}
\lv{\usepackage{authblk}}
\usepackage{amsmath}
\lv{\usepackage{amsthm}}
\usepackage{xspace}
\usepackage{mwe}\usepackage[hypcap=false]{caption} %
\usepackage[colorlinks=true,allcolors=blue]{hyperref}

\usepackage[color=Olive]{todonotes}

\lv{%
\newtheorem{lemma}{Lemma}[section]
\newtheorem{theorem}[lemma]{Theorem}
                              
\newtheorem{proposition}[lemma]{Proposition}
}

\usepackage[absolute]{textpos}
\title{%
  On 3-Coloring of $(2P_4,C_5)$-Free Graphs\thanks{
    V.~Jelínek was supported by project 18-19158S of the Czech Science Foundation.
    T.~Klimo\v sov\'a is supported by the grant no.~19-04113Y of the Czech Science Foundation (GA\v{C}R) and the Center for Foundations of Modern Computer Science (Charles Univ. project UNCE/SCI/004).
    T.~Masařík received funding from the European Research Council (ERC) under the European Union's Horizon 2020
    research and innovation programme Grant Agreement 714704.
    He completed a part of this work while he was a postdoc at Simon Fraser University in Canada.
    J.~Novotná and A.~Pokorná were supported by SVV-2017-260452 and GAUK 1277018.
    }\lv{~\thanks{
An extended abstract of this paper has been accepted to the proceedings of International Workshop
on Graph-Theoretic Concepts in Computer Science (WG) 2021.
}}\sv{~\thanks{A preprint of the full version of this paper is available from arXiv~\cite{2P4preprint}.}}
}

\lv{%
\author[1]{Vít Jelínek}
\author[1]{Tereza Klimošová}
\author[1,2,3]{Tomáš Masařík}
\author[1,2]{\\Jana Novotná}
\author[1]{Aneta Pokorná}

\affil[1]{Faculty of Mathematics and Physics, Charles University, Prague, Czech Republic}
\affil[2]{University of Warsaw, Poland}
\affil[3]{Simon Fraser University, Burnaby, Canada}
\affil[ ]{\texttt{\{jelinek, pokorna\}@iuuk.mff.cuni.cz}}
\affil[ ]{\texttt{\{tereza,masarik,janca\}@kam.mff.cuni.cz}}
\date{}
}

\sv{
\author{Vít Jelínek\inst{1}%
  \and
Tereza Klimošová\inst{1}%
\and
Tomáš Masařík\inst{1,2,3}%
\and\\
Jana Novotná\inst{1,2}%
\and
Aneta Pokorná\inst{1}%
}

\authorrunning{V.~Jelínek et al.}

\institute{Faculty of Mathematics and Physics, Charles University, Prague, Czech Republic\and
University of Warsaw, Poland\and
Simon Fraser University, Burnaby, Canada\\
\email{\{jelinek, pokorna\}@iuuk.mff.cuni.cz}\\
\email{\{tereza,masarik,janca\}@kam.mff.cuni.cz}
}
}

\begin{document}

\maketitle
\lv{
\begin{textblock}{20}(0, 14.0)
\includegraphics[width=40px]{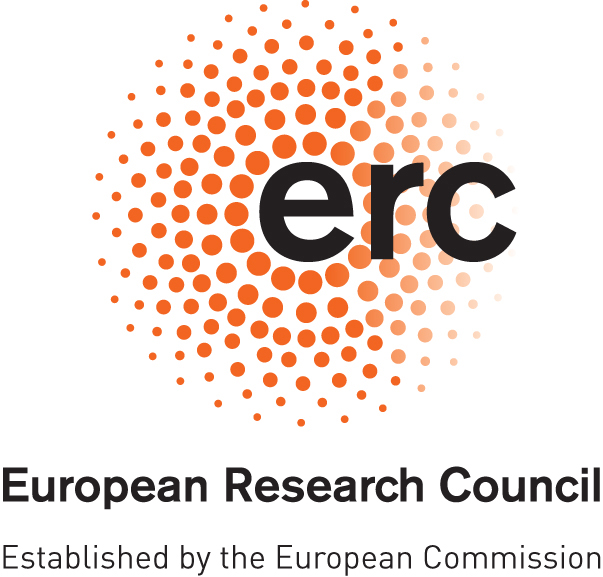}%
\end{textblock}
\begin{textblock}{20}(0, 14.9)
\includegraphics[width=40px]{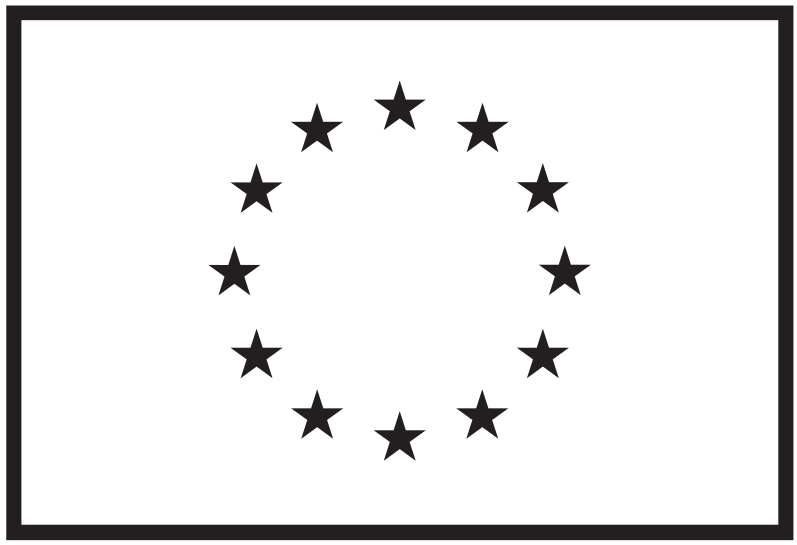}%
\end{textblock}
}

\begin{abstract}
The 3-coloring of hereditary graph classes has been a deeply-researched
problem in the last decade. A hereditary graph class is characterized by a
(possibly infinite) list of minimal forbidden induced subgraphs
$H_1,H_2,\ldots$; the graphs in the class are called
$(H_1,H_2,\ldots)$-free.
The complexity of 3-coloring is far from being
understood, even for classes defined by a few small forbidden induced
subgraphs.
For $H$-free graphs, the
complexity is settled for any $H$ on up to seven vertices.
There are
only two unsolved cases on eight vertices, namely $2P_4$ and $P_8$. For
$P_8$-free graphs, some partial results are known, but to the best of our
knowledge, $2P_4$-free graphs have not been explored yet. In this paper,
we show that the 3-coloring problem is polynomial-time solvable on
 $(2P_4,C_5)$-free graphs.
\end{abstract}
\sv{%
}

\section{Introduction}

Graph coloring is a notoriously known and well-studied concept in both graph theory and theoretical computer science. 
A \emph{$k$-coloring} of a graph $G=(V,E)$ is defined as a mapping $c:V\to \{1,\ldots,k\}$ which is 
\emph{proper}, i.e., it assigns distinct colors to $u,v\in V$ if $uv\in E$.
The \textsc{$k$-coloring} problem asks whether a given graph admits a $k$-coloring.
For any $k\ge 3$, the $k$-coloring is a well-known NP-complete problem~\cite{Karp72}.
We also define a more general \emph{list-$k$-coloring} where each vertex $v$ has a list $P(v)$ of 
allowed colors such that $P(v)\subseteq \{1,\ldots,k\}$.
In that case, the coloring function $c$, in addition to being proper, has to respect the lists, 
that is, $c(v)\in P(v)$ for every vertex~$v$.

A graph class is \emph{hereditary} if it is closed under vertex deletion.
It follows that a  graph class  $\mathcal{G}$  is hereditary if and only if $\mathcal G$  can be characterized by a unique (not necessarily finite) set $\mathcal H_{\mathcal G}$ of minimal forbidden induced subgraphs.
Special attention was given to hereditary graph classes where $\mathcal H_{\mathcal G}$ contains only one or only a very few elements.
In such cases, when $\{H\}=\mathcal H_{\mathcal G}$, or $\{H_1,H_2,\ldots\}=\mathcal H_{\mathcal G}$, we say that $G\in\mathcal G$ is \emph{$H$-free}, or \emph{$(H_1,H_2,\ldots)$-free}, respectively.
We let $P_t$ denote the path on $t$ vertices, and $C_\ell$ the cycle on $\ell$ vertices.
We let $\overline{H}$ denote the complement of a graph~$H$. For two graphs $H_1$ and $H_2$, we let 
$H_1+H_2$ denote their disjoint union\sv{. }\lv{, and w}\sv{W}e write $kH$ for the disjoint union of $k$ copies of a 
graph~$H$.

In recent years, a lot of attention has been paid to determining the complexity of $k$-coloring of 
$H$-free graphs.
Classical results imply that for every $k\ge 3$, $k$-coloring of $H$-free graphs is NP-complete if 
$H$ contains a cycle~\cite{EHK98} or an induced claw~\cite{Holyer81,LevenGalil83}. 
Hence, it remains to consider the cases where $H$ is a \emph{linear forest}, i.e., a disjoint union of paths.
The situation around complexity of (list) $k$-coloring on $P_t$-free graphs where $k\ge 4$ has been resolved completely.
The cases $k=4,t\ge 7$ and $k\ge 5,t\ge 6$ are NP-complete~\cite{Huang16} while cases for $k\ge 
1,t=5$ are polynomial-time solvable~\cite{HKLSS08}.
In fact, $k$-coloring is polynomial-time solvable on $sP_1+P_5$-free graphs for any $s\ge 0$~\cite{CGKP13}.
The borderline case where $k=4,t=6$ has been settled recently.
There the $4$-coloring problem (even the precoloring extension problem with 4 colors) is 
polynomial-time solvable~\cite{SCZ19} while the list 4-coloring problem is 
NP-complete~\cite{GPS14_IC}.
\lv{In 2021, Hajebi, Li, and Spirkl show that 5-coloring on $2P_4$-free graphs is NP-complete~\cite{5colorSpirkl}.}

Now, we move our focus towards the complexity of the 3-coloring problem, which was less well understood, in spite of the amount of the research interest it received  in the past years. 
However, a considerable progress has been made in 2020; a quasi-polynomial algorithm running in 
time $n^{O(\log^2(n))}$ on $n$-vertex $P_t$-free graphs ($t$ is a constant) was shown by Pilipczuk 
et al.~\cite{PPR20}\sv{ (extending results in~\cite{GL20})}\lv{, extending a breakthrough of Gartland and Lokshtanov~\cite{GL20}}.
In the realms of polynomiality, Bonomo et al.~\cite{BCMSZ17}
found a polynomial-time algorithm for $P_7$-free graphs. Klimošová et al.~\cite{KMMNPS20} completed 
the classification of 3-coloring of $H$-free graphs, for any $H$ on up to $7$ vertices.
These results were subsequently extended to $P_6+rP_3$-free graphs, for any $r\geq 0$~\cite{CHSZ20}.
There are only two remaining graphs on at most $8$ vertices, namely $P_8$ and $2P_4$, for 
which the complexity of 3-coloring is still unknown.
\lv{%

}%
Algorithms for subclasses of $P_t$-free graphs which avoid one or more additional  induced subgraphs, usually cycles, have been studied. They  might be a first step in the attempt to settle the case of $P_t$-free graphs. This turned out to be the case for 3-coloring of $P_7$-free graphs (as can be seen from preprints \cite{BSS14,CMZ14,CMZ15} leading to~\cite{BCMSZ17}) and 4-coloring of  $P_6$-free graphs~\cite{CMSZ17}.
\lv{%

}%
Note that the problem of $4$-coloring is NP-complete even when some $(P_t,C_\ell)$-free graphs are considered when $t\geq 7$.
Hell and Huang~\cite{HellHuang17} and Huang et al.~\cite{HJP15} settled many NP-complete cases of this type. 
These results, in combination 
with the polynomiality of $P_6$-free case, leave open only the following cases: $(P_7,C_7)$-, 
$(P_8,C_7)$-, and $(P_t,C_3)$-free graphs, for $7\le t\le 21$. 

Chudnovsky and Stacho~\cite{ChudnovskyStacho18} studied the problem of $3$-coloring of $P_8$-free graphs which additionally avoid induced cycles of two distinct lengths; specifically, they consider graphs that are
\lv{$(P_8,C_3,C_4)$-free, $(P_8,C_3,C_5)$-free, and $(P_8,C_4,C_5)$-free.}%
\sv{$(P_8,C_3,C_4)$-, $(P_8,C_3,C_5)$-, and $(P_8,C_4,C_5)$-free.}
For the first two cases, they show that all such graphs are 3-colorable.
For the last one, they provide a complete list of \emph{$3$-critical graphs}, i.e., the graphs with 
no $3$-coloring such that all their proper induced subgraphs are 3-colorable.
Independently, using a computer search, Goedgebeur and Schaudt~\cite{GoedgebeurSchaudt17} showed 
that there are only finitely many $3$-critical ($P_8, C_4$)-free graphs.
In fact, 3-coloring is polynomial-time solvable on ($P_t,C_4$)-free graphs for any $t\geq 1$~\cite{GPS14_DAM}.

The situation concerning $2P_4$ or $P_8$ is still far from being determined when two forbidden 
induced subgraphs are considered; in particular, it is not known whether $(P_8,C_3)$-\lv{free}, 
$(P_8,C_5)$-\lv{free}, $(2P_4,C_3)$-\lv{free}, or $(2P_4,C_5)$-free graphs can be 3-colored in polynomial 
time\footnote{First two cases were explicitly mentioned as open in~\cite{GJPS16} and \cite{RojasStein20}, the 
latter two cases are open to the best of our knowledge.}.
This is in contrast with the algorithm for $(P_7,C_3)$-free graphs~\cite{BCGMSSZ14} which is considerably simpler than the one for $P_7$-free graphs~\cite{BCMSZ17}.
Recently, Rojas and Stein~\cite{RojasStein20} approached the problem  by showing that for any odd $t\ge 9$, there exists a polynomial-time algorithm that solves the 3-coloring problem in $P_t$-free graphs of odd girth at least $t-2$.
In particular, their result implies that 3-coloring is polynomial-time solvable for $(P_9,C_3,C_5)$-free graphs.

Freshly, a similar question was resolved in the case where, instead of a cycle, a 1-subdivision of $K_{1,s}$ (a star with $s$ leaves), denoted as $SDK_{1,s}$, is considered.
Chudnovsky, Spirkl, and Zhong have shown that the class of $(SDK_{1,s},P_t)$-free graphs is 
list-3-colorable in polynomial time for any $s,t\geq 1$~\cite{CSZ20}.
For other related results and history of the problem, please consult a recent survey~\cite{GJPS16}.

In this paper, we resolve one of the remaining open problems mentioned above, which considers $2P_4$-free graphs, as we will describe a polynomial-time algorithm for 3-coloring of $(2P_4,C_5)$-free graphs.
To the best of our knowledge, this is a first attempt to attack the 3-coloring of $2P_4$-free graphs.

\begin{theorem}\label{thm:main}
  The 3-coloring problem is polynomial-time solvable on $(2P_4,C_5)$-free graphs.
\end{theorem}

To prove our result, we will make use of some relatively standard techniques.
Let $\omega(G)$ be the size of the largest clique of graph $G$.
We use a seminal result of Grötschel, Lovász, and Schrijver~\cite{GLS84} that shows the $k$-coloring 
problem on \emph{perfect graphs}, i.e., graphs where each induced subgraph $G'$ is 
$\omega(G')$-colorable, can be solved in polynomial time.
Perfect graphs are characterized by the strong perfect graph theorem~\cite{CRST06} as the graphs that have neither odd-length induced cycles nor complement of odd-length induced cycles on at least five vertices.

As $K_4$ and $\overline{C_7}$ graphs are not 3-colorable, we can assume that our graph is 
$(2P_4,C_5,\overline{C_7},K_4)$-free.
As $K_4\subseteq\overline{C_\ell}$ whenever $\ell\ge8$ and $2P_4\subseteq C_\ell$ whenever $\ell\geq 
10$, it follows that either the graph is perfect, or it contains $C_7$ or~$C_9$.
In the first case, we are done by the aforementioned polynomial-time algorithm.
For the latter cases, we divide the analysis into two further subcases.
First, we suppose that the graph is $(2P_4,C_5, C_7,\overline{C_7},K_4)$-free and therefore it must 
contain~$C_9$.
\lv{We analyze this case in Subsection~\ref{s:c7free}.
Second, we suppose that graph contains $C_7$ and we analyze this case in Subsection~\ref{s:withc7}.}
\sv{We first analyze this case then we suppose that graph contains $C_7$ for the rest of the proof.}
\lv{%

}%
We will exploit the fact that once we find an induced $P_4$, the vertices that are not adjacent to it must induce a $P_4$-free graph (also known as \emph{cograph}).
Such graphs were among the first $H$-free graphs studied, and have many nice properties, e.g., any 
greedy coloring gives a proper coloring using the least number of colors~\cite{CS79}.
We will make use of a slightly stronger statement that handles the list-$3$-coloring problem.

\begin{theorem}[\cite{GJPS16}]\label{th:cographs}
The list-3-coloring problem on $P_4$-free graphs can be solved in polynomial time.
\end{theorem}

The 3-coloring algorithm that we develop to prove Theorem~\ref{thm:main} cannot be directly extended 
to solve the more general list-3-coloring problem, since it uses the 3-coloring algorithm for perfect 
graphs to deal with graphs avoiding $C_7$ and $C_9$. However, apart from this one case, the algorithm 
works with the more general setting of list-3-coloring. In fact, we use reductions of lists as one of 
our base techniques. After several branching steps with polynomially many branches and suitable 
structural reductions of the original graph $G$, the algorithm will transform a 3-coloring instance 
of a $(2P_4,C_5)$-free graph $G$ to a set of polynomially many heavily structured list-3-coloring 
instances. These structured instances can then be encoded by a 2-SAT formula, whose satisfiability 
is solvable in linear time~\cite{K67}.

\section{Proof of Theorem~\ref{thm:main}}

We are given a $(2P_4,C_5)$-free graph $G=(V,E)$, and our goal is to determine whether it is 
3-colorable. We will present an algorithm that solves this problem in polynomial time. The algorithm 
begins by checking that the graph is $\overline{C_7}$-free, and that the neighborhood of each vertex 
induces a bipartite graph, rejecting the instance if the check fails. Note that this check ensures, 
in particular, that $G$ is $K_4$-free.

The algorithm then partitions the graph into connected components, solving the 3-coloring problem 
for each component separately. From now on, we assume that the graph $G=(V,E)$ is connected, 
$\overline{C_7}$-free, and each of its vertices has a bipartite neighborhood.
\lv{%

}%
The basic idea of the algorithm is to choose an initial subgraph $N_0$ of bounded size, try all 
possible proper 3-colorings of $N_0$, and analyze how the precoloring of $N_0$ affects the possible 
colorings of the remaining vertices. 
\lv{%

}%
We let $N_1$ denote the vertices in $V\setminus N_0$ which are adjacent to at least one vertex of 
$N_0$, and we let $N_2$ be the set $V\setminus (N_0\cup N_1)$. We will use the notation $N(x)$ for 
the set of neighbors of $x$ in $G$, and $N_i(x)$ for $N_i\cap N(x)$.

Our algorithm will iteratively color the vertices of~$G$. We will assume that throughout the 
algorithm, each vertex $v$ has a list $P(v)\subseteq\{1,2,3\}$ of \emph{available colors}. We call 
$P(v)$ the \emph{palette of $v$}. The goal is then to find a proper coloring of $G$ in which each 
vertex is colored by one of its available colors. The problem of deciding the existence of such 
coloring is known as the \emph{list-3-coloring problem}, and is a generalization of the 3-coloring 
problem.
\lv{%

}%
Whenever a vertex $x$ of $G$ is colored by a color $c$ in the course of the algorithm, we 
immediately remove $c$ from the palette of $x$'s neighbors. Additionally, if the vertex $x$ is not 
in $N_0$, it is then deleted. The vertices in $N_0$ are kept in $G$ even after they are colored. We 
then update the list-3-coloring instance using the following \emph{basic reductions}:
\begin{itemize}
\item If a vertex $y$ has only one color $c'$ left in $P(y)$, we color it by the color $c'$ and 
remove $c'$ from the palettes of its neighbors. If $y\not \in N_0$, we then delete~$y$.
\item If $P(y)$ is empty for a vertex $y$, the instance of list-3-coloring is rejected.
\item If, for a vertex $y\not\in N_0$, the size of $P(y)$ is greater 
than the degree of $y$, we delete~$y$.
\item \emph{Diamond consistency rule}:
  If $y$ and $y'$ are a pair of nonadjacent vertices such that $P(y)\neq P(y')$, and if 
$N(y)\cap N(y')$ is not an independent set, 
then any valid 3-coloring of $G$ must assign the same 
color to $y$ and $y'$; we therefore replace both $P(y)$ and $P(y')$ with $P(y)\cap P(y')$. 
\item \emph{Neighborhood domination rule}:
  If $y$ and $y'$ are a pair of nonadjacent vertices such that $N(y)\subseteq N(y')$ and 
$P(y')\subseteq P(y)$, and if $y$ is not in $N_0$, we delete~$y$. 
\item If $G$ has a connected component in which every vertex has at most two available colors, we 
determine 
whether the component is colorable by reducing the problem to an instance of 2-SAT. If the component 
can be colored, we remove it from $G$ and continue, otherwise we reject the whole instance.
\item If a connected component of $G$ is $P_4$-free, we solve the list-3-coloring problem for this component by Theorem~\ref{th:cographs}.
If the component is colorable we remove it, 
otherwise we reject the whole instance~$G$.
\end{itemize}

It is clear that the rules are correct in the sense that the instance of list-3-coloring produced by 
a basic reduction is list-3-colorable if and only if the original instance was list-3-colorable. It 
is also clear that we may determine in polynomial time whether an instance of list-3-coloring (with 
fixed $N_0$) permits an application of a basic reduction, and perform the basic reduction, if 
available. Throughout the algorithm, we apply the basic reductions greedily as 
long as possible, until we reach a situation where none of them is applicable. 

The basic reductions by themselves are not sufficient to solve the 3-coloring problem for~$G$. Our 
algorithm will sometimes also need to perform branching, i.e., explore several alternative ways to 
color a vertex or a set of vertices. Formally, this means that the algorithm reduces a given instance 
$G$ of list-3-coloring to an equivalent set of instances $\{G_1,\dotsc,G_k\}$; here saying that a 
list-3-coloring instance $G$ is \emph{equivalent} to a set $\{G_1,\dotsc,G_k\}$ of instances means 
that $G$ has a solution if and only if at least one of $G_1,\dotsc,G_k$ has a solution. 

In the beginning of the algorithm, we attach to each vertex $v$ of $G$ the list $P(v)=\{1,2,3\}$ of 
available colors, thereby formally transforming it to an instance of list-3-coloring. The algorithm 
will then try all possible proper 3-colorings of~$N_0$, and for each such coloring, apply basic 
reductions as long as any basic reduction is applicable. If this fails to color all the vertices, 
more complicated reduction steps and further branching will be performed, to be described later.
\lv{%

}%
Overall, the algorithm will ensure that the initial instance $G$ is eventually reduced to a set of 
at most polynomially many smaller instances, each of which can be transformed to an equivalent 
instance of 2-SAT, which then can be solved efficiently.

\lv{\subsection{The \texorpdfstring{$C_7$}{C\_7}-free case}\label{s:c7free}}
\lv{
Our choice of $N_0$ will depend on the structure of $G$. More precisely, if $G$ contains an induced 
copy of $C_7$, we will choose one such copy as~$N_0$. This is by far the most challenging case, and 
we return to it later.
}

The case when $G$ is $C_7$-free can be handled in a simple way\lv{, as we now show}. 

\begin{proposition}\label{pro-c7free} \sv{($\clubsuit$)\footnote{
      Due to space limitations, we defer some proofs to the full version of our paper~\cite{2P4preprint}.
We mark the respective statements by $(\clubsuit)$.
    }%
  }
The 3-coloring problem for a $(2P_4,C_5,C_7)$-free graph $G$ can be solved in polynomial time.
\end{proposition}
\toappendix{
  \lv{\begin{proof}}
    \sv{\begin{proof}[of Proposition~\ref{pro-c7free}]}
Recall that we assume that $G$ is $K_4$-free and $\overline{C_7}$-free; otherwise $G$ would clearly 
not be 3-colorable. Note that $K_4$-freeness implies that $G$ is $\overline{C_k}$-free for every 
$k\ge 8$, and $2P_4$-freeness implies that $G$ is $C_k$-free for every $k\ge 10$. 

If $G$ is also $C_9$-free, then it is perfect by the strong perfect graph theorem, and since it is 
$K_4$-free, it is 3-colorable. Assume then that $G$ contains an induced copy of~$C_9$.
Fix $N_0$ to be an induced copy of $C_9$ in~$G$, and define $N_1$ and $N_2$ accordingly. We will 
show that for any proper coloring of $N_0$, the basic reductions can solve the resulting 
list-3-coloring problem.

Fix a 3-coloring of~$N_0$, and apply the basic reductions, until none of them is applicable. We 
claim that this solves the instance completely, i.e., we either color the whole graph, or determine 
that no coloring exists. For contradiction, suppose that we reached a situation when $G$ still 
contains uncolored vertices, but no basic reduction is applicable. 

It follows that $G$ contains a vertex with three available colors, and this vertex necessarily 
belongs to $N_2$. In particular, $N_2$ is nonempty, and therefore we may find in $G$ two adjacent 
vertices $x,y$ with $x\in N_1$ and $y\in N_2$. Recall that $N_0(x)$ is the set of vertices of $N_0$ 
adjacent to~$x$. The vertices of $N_0(x)$ partition the cycle $N_0$ into edge-disjoint arcs, and at 
least one of these arcs has an odd number of edges. Let $A$ be such an arc of odd length. 

If $A$ has length 1, then $x$ is adjacent to two adjacent vertices of $N_0$, hence the color of $x$ 
is uniquely determined by the coloring of $N_0$ and $x$ should have been deleted. If $A$ has 
length 3 or 5, then $A\cup\{x\}$ induces a copy of $C_5$ or $C_7$, respectively, which 
is impossible. Thus, $A$ has length 7 or 9. In such case, we find a copy of $2P_4$ in 
$G$, where one $P_4$ consists of $y$, $x$, a vertex $z\in N_0(x)$, and a vertex $w\in A$ adjacent to 
$z$, while the other $P_4$ is formed by taking four consecutive internal vertices of $A$, each of 
which is at distance at least two from $z$ and $w$\sv{.}\lv{; see Figure~\ref{fig:2P4C5C7}.} In all cases we 
get a contradiction.
\sv{\qed}
\end{proof}

\begin{figure}
\centering
\includegraphics[width=0.4\textwidth]{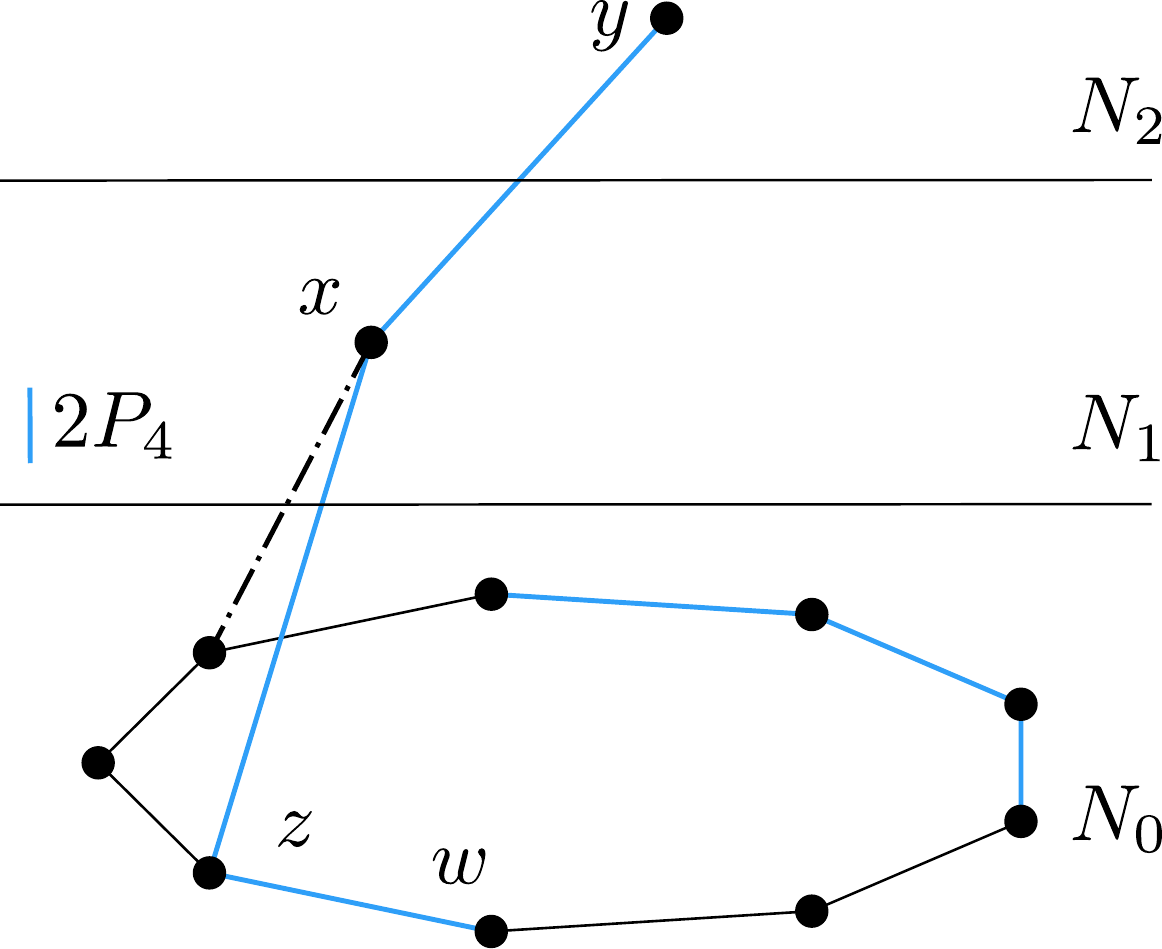}
\captionof{figure}{Picture showing the induced $2P_4$ in the case of $G$ being $C_7$-free.
If the dash-and-dotted edge is present, $A$ has length 7, otherwise $A$ has length 9.}
\label{fig:2P4C5C7}
\end{figure}
}

From now on, we assume that the graph $G$ contains an induced $C_7$. We choose one such $C_7$ as
$N_0$, and define $N_1$ and $N_2$ accordingly.

\lv{\subsection{More complicated reductions}}
\sv{\medskip\noindent\textbf{More complicated reductions.~~}}
Apart from the basic reductions described previously, which we will apply whenever opportunity 
arises, we will also use more complicated reductions, to be applied in specific situations.

\lv{\paragraph*{Cut reduction.}}
\sv{\smallskip\noindent\textit{Cut reduction.~~}}
Suppose $G=(V,E)$ is a connected instance of list-3-coloring. Let 
$X\subseteq V$ be a vertex cut of $G$, let $C$ be a union of one or more connected components of 
$G-X$, and let $C_X$ be the subgraph of $G$ induced by $C\cup X$. 
Suppose further that the following conditions hold.
\begin{itemize}
\item $C$ has at least two vertices.
\item $X$ is an independent set in $G$.
\item All the vertices in $X$ have the same palette, which has size 2.
\item For any two vertices $x,x'$ in $X$, we have $N(x)\cap C= N(x')\cap C$.
\item The graph $C_X$ is $P_4$-free.
\end{itemize}
Assume without loss of generality that all the vertices of $X$ have palette equal to $\{1,2\}$. Let 
us say that a coloring $c\colon X\to\{1,2\}$ of $X$ is \emph{feasible for $C$}, if it can be 
extended into a proper 3-coloring of the list-3-coloring instance~$C_X$. Note that the feasibility 
of a given coloring can be determined in polynomial time by Theorem~\ref{th:cographs}, because $C_X$ is 
a cograph.

We distinguish three types of possible colorings of $X$: the \emph{all-1} coloring colors all the 
vertices of $X$ by the color 1, the \emph{all-2} coloring colors all the vertices of $X$ by color 2, 
and a \emph{mixed} coloring is a coloring that uses both available colors on~$X$. Observe that if 
$X$ admits at least one mixed coloring feasible for $C$, then every (not necessarily mixed) coloring 
of $X$ by colors 1 and 2 is feasible for~$C$. This is because when we extend a mixed coloring of $X$ 
to a coloring of $C_X$, all the vertices $y \in C$ must receive the color~$3$. If such a coloring 
of $C$ exists, we can combine it with any coloring of~$X$ by colors 1 and~2.

The \emph{cut reduction} of $X$ and $C$ is an operation which reduces $G$ to a smaller, equivalent 
list-3-coloring instance, determined as follows. We choose an arbitrary mixed coloring $c$ of $X$, 
and check whether it is feasible for~$C$. If it is feasible, we reduce the instance $G$ to $G-C$, 
leaving the palettes of the remaining vertices unchanged. The new instance is equivalent to the 
original one, since any proper list-3-coloring of $G-C$ can be extended to a coloring of $G$, 
because all the colorings of $X$ are feasible for~$C$. 

If the mixed coloring $c$ is not feasible for $C$, we know that no mixed coloring is feasible. We 
then test the all-1 and the all-2 coloring for feasibility. If both are feasible, we reduce the 
instance $G$ by replacing $C$ with a single new vertex $v$, with palette $P(v)=\{1,2\}$, and 
connecting $v$ to all the vertices of~$X$. Note that the reduced instance is an induced subgraph of 
the original one. It is easy to see that the reduced instance is equivalent to the original one.

If only one coloring of $X$ is feasible for $C$, we delete $C$, color the vertices of $X$ using the 
unique feasible coloring, and delete the corresponding color from the palettes of the neighbors 
of~$X$ in~$G-C$. If no coloring of $X$ is feasible for $C$, we declare that $G$ is not 
list-3-colorable.

\lv{\paragraph{Neighborhood collapse.}}
\sv{\smallskip\noindent\textit{Neighborhood collapse.~~}}
Let $G$ be an instance of list-3-coloring, and let $v$ be a 
vertex of~$G$. Suppose that $N(v)$ induces in $G$ a connected bipartite graph with nonempty partite 
classes $X$ and~$Y$. Suppose furthermore that all the vertices of $X$ have the same palette $P_X$, 
and all the vertices in $Y$ have the same palette~$P_Y$. The \emph{neighborhood collapse} of the 
vertex $v$ is the operation that replaces $X$ and $Y$ by a pair of new vertices $x$ and $y$, 
adjacent to each other and to $v$, with the property that any vertex of $G-Y$ adjacent to at 
least one vertex in $X$ will be made adjacent to $x$, and similarly every vertex adjacent to $Y$ in 
$G-X$ will be adjacent to~$y$. We then set $P(x)=P_X$ and $P(y)=P_Y$. 
\lv{Informally speaking, we have collapsed the vertices in $X$ to a single vertex $x$, and similarly for $Y$ and~$y$.}

It is clear that the collapsed instance is equivalent to the original one. However, since the new 
instance is not necessarily an induced subgraph of the original one, it might happen, e.g., that a 
collapse performed in a $C_5$-free graph will introduce a copy of $C_5$ in the collapsed instance. 
In our algorithm, we will only perform collapses at a stage when $C_5$-freeness will no longer be 
needed. 

\sv{\stepcounter{lemma}}
On the other hand, $2P_4$-freeness is preserved by collapses, as we now show.
\begin{lemma}\label{lem:2P4freeness}\sv{($\clubsuit$)}
Let $G$ be a $2P_4$-free instance of list-3-coloring in which a neighborhood collapse of a vertex 
$v$ may be performed, and let $G^*$ be the graph obtained by the collapse. Then $G^*$ is $2P_4$-free.
\end{lemma}
\toappendix{
\sv{\begin{proof}[of Lemma~\ref{lem:2P4freeness}]}
\lv{\begin{proof}}
Suppose $G^*$ contains an induced $2P_4$, and let $P$ and $Q$ be the two nonadjacent copies 
of~$P_4$. Let $x$ and $y$ be the two vertices obtained by collapsing sets $X$ and $Y$, as in the 
definition of neighborhood collapse. Without loss of generality, $P$ contains the vertex~$x$. It 
follows that $Q$ contains none of $x$, $y$ or~$v$, and in particular, $Q$ is also a $P_4$ 
in~$G$. 

If the path $P$ contains the edge $xy$, we may `lift' $P$ into the graph $G$ by replacing the 
vertices $x$ and $y$ by appropriate vertices $x'\in X$ and $y'\in Y$, and by replacing the edge $xy$ 
by a shortest path from $x'$ to $y'$ in $N(v)$. This transforms $P$ into an induced path $P'$ in 
$G$ on at least four vertices which is nonadjacent to~$Q$. Thus, $G$ also contains a $2P_4$.

Suppose now that $P$ does not contain the edge $xy$, and therefore $y$ is not in~$P$. If $x$ is the 
end-vertex of $P$, say $P=xw_1w_2w_3$, we easily obtain a $2P_4$ in $G$ by simply replacing $x$ by a 
vertex $x'\in X$ adjacent to $w_1$ in~$G$. Suppose then that $x$ is an internal vertex of $P$, say 
$P=w_1x w_2w_3$. Since we know that $P$ does not contain $y$, we may replace the vertex $w_1$ with 
$v$ in $P$, knowing that $vxw_2w_3$ is also an induced $P_4$ in $G^*$ nonadjacent to~$Q$. By 
replacing the vertex $x$ by a vertex $x'\in X$ adjacent to $w_2$, we obtain the induced path 
$vx'w_2w_3$ in $G$ which forms a $2P_4$ together with~$Q$.
\sv{\qed}
\end{proof}
}

\lv{\subsection{Graphs containing \texorpdfstring{$C_7$}{C\_7}}\label{s:withc7}}
\sv{\noindent\textbf{Graphs containing \texorpdfstring{$C_7$}{C\_7}.~~}\label{s:withc7}}
We now turn to the most complicated part of our coloring algorithm, which solves the 3-coloring 
problem for a $(2P_4,C_5)$-free graph $G$ that contains an induced $C_7$. We let $N_0$ be an 
induced copy of $C_7$ in this graph, and define $N_1$ and $N_2$ accordingly.

We let $v_1,v_2,\dotsc,v_7$ denote the vertices of $N_0$, in the order in which they appear on the 
cycle~$N_0$. We evaluate their indices modulo $7$, so that, e.g., $v_8=v_1$.

Fix a proper coloring of $N_0$, and apply the basic reductions to $G$ until no basic reduction is 
applicable. We now analyze the structure of $G$ at this stage of the algorithm. We again let 
$N_0(x)$ denote the set of neighbors of $x$ in~$N_0$.

\begin{lemma}\label{lem-n0}\sv{($\clubsuit$)}
After fixing the coloring of $N_0$ and applying all available basic reductions, the graph $G$ has 
the following properties.
\begin{itemize}
\item Each vertex $x$ of $N_1$ satisfies either $N_0(x)=\{v_i\}$ for some $i$, or 
$N_0(x)=\{v_i,v_{i+2}\}$ for some~$i$.
\item Each induced copy of $P_4$ in $G$ has at most two vertices in $N_2$.
\item $G$ is connected.
\end{itemize}
\end{lemma}

\toappendix{
  \lv{\begin{proof}}
    \sv{\begin{proof}[of Lemma~\ref{lem-n0}]}
To prove the first part, use the vertices of $N_0(x)$ to partition the cycle $N_0$ into 
edge-disjoint arcs. Note that none of these arcs has length 1, since then $x$ would be adjacent to 
two vertices of distinct colors, and it would have been colored and deleted. Also, none of these 
arcs can have length 3, since such an arc together with the vertex $x$ would induce a $C_5$ in 
$G$, contradicting $C_5$-freeness. 

On the other hand, at least one of the arcs formed by $N_0(x)$ must have odd length. Thus, there is 
either an arc of length 7, implying $N_0(x)=\{v_i\}$, or there is an arc of length 5, implying 
$N_0(x)=\{v_i,v_{i+2}\}$ for some~$i$. This proves the first part of the lemma.

\bigskip
\begin{minipage}{0.40\textwidth}
\centering
\includegraphics[width=0.95\textwidth]{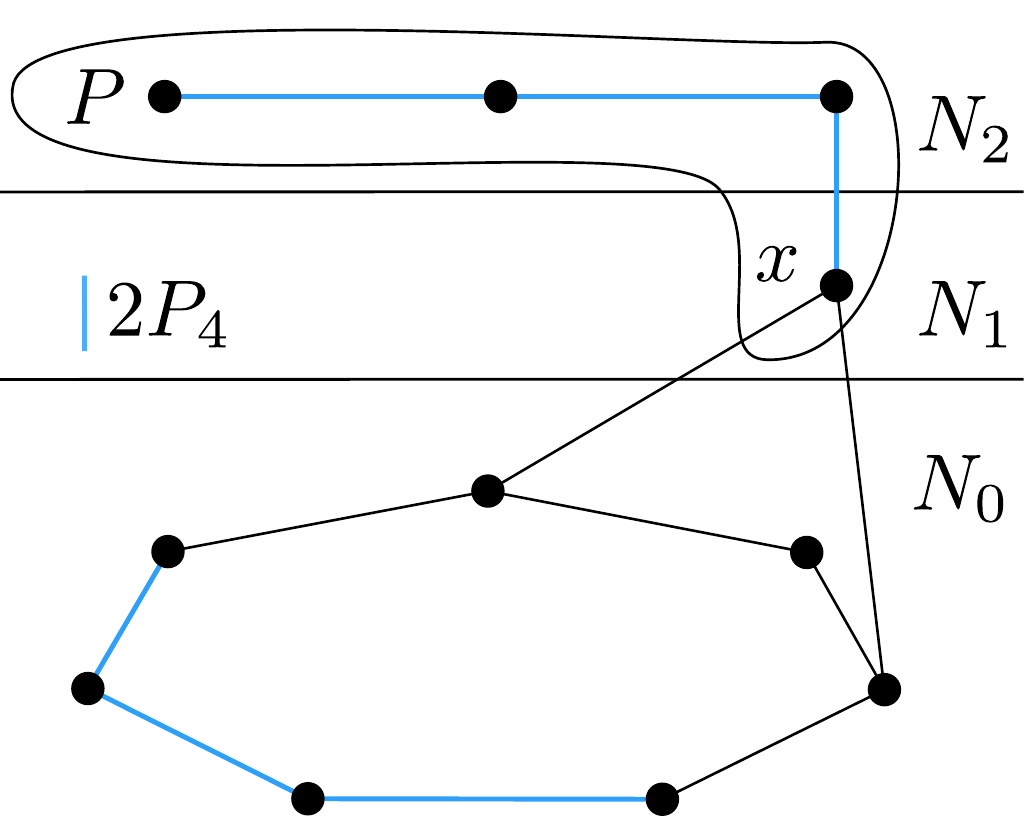}
\captionof{figure}{Finding an induced $2P_4$, assuming $P$ is an induced 
$P_4$ with exactly three vertices in $N_2$. Note that $P$ can look differently, but always 
contains~$x$.}
\label{fig:3-vertices-N2}
\end{minipage}
~
\begin{minipage}{0.50\textwidth}
\centering
\includegraphics[width=0.95\textwidth]{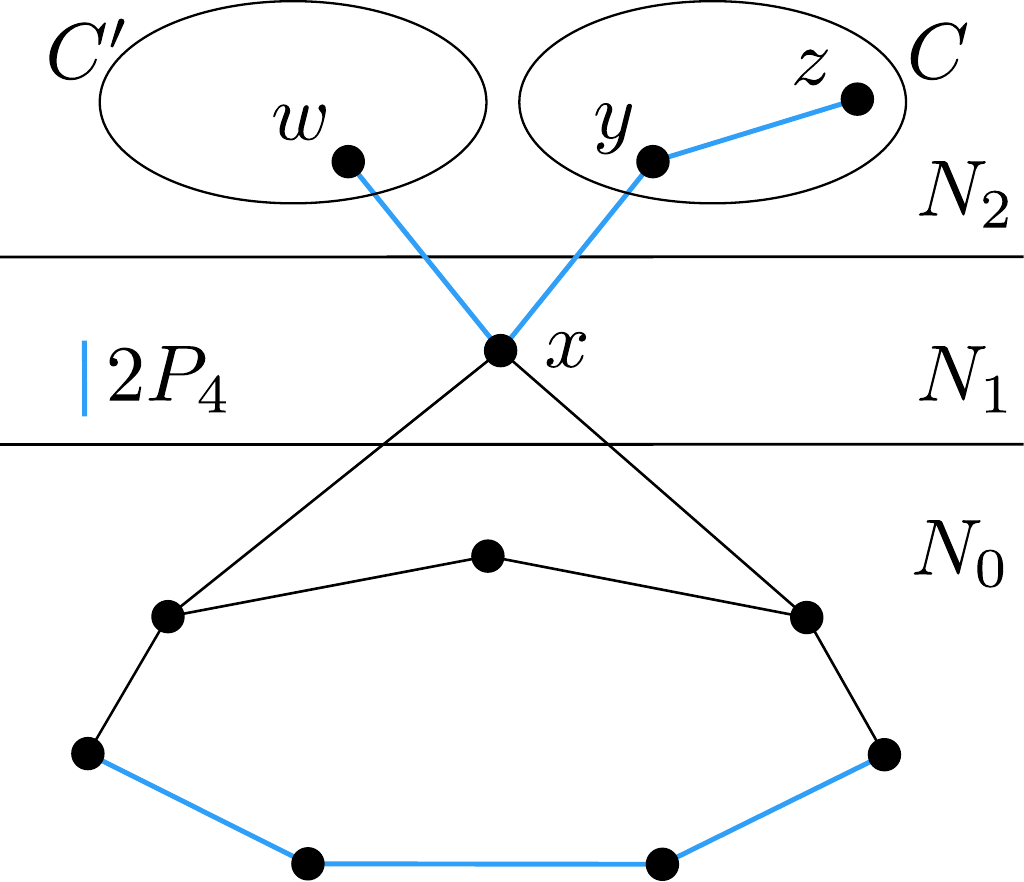}
\captionof{figure}{Vertex $x \in N_1$ being a partial neighbor of a top component $C$ and neighboring another top component $C'$ leads to an induced $2P_4$.}
\label{fig:N0x2}
\end{minipage}
\bigskip

To prove the second part, assume that $P$ is an induced copy of $P_4$ in~$G$ with at least three 
vertices in $N_2$. If $P$ is fully contained in $N_2$, then $P$ forms a $2P_4$ together with any 
$P_4$ contained in $N_0$. Suppose that  $P\setminus N_2$ consists of a single vertex~$x$, as in Figure~\ref{fig:3-vertices-N2}. Necessarily 
$x$ is in $N_1$, and by the first part of the lemma, $N_0\setminus N_0(x)$ contains an induced $P_4$ 
which forms an induced $2P_4$ with~$P$.

To prove the last part of the lemma, note that $N_0$ is connected and therefore contained in a 
single component of $G$, and if $G$ contained another connected component, then this other component 
would necessarily be $P_4$-free and would be colored by a basic reduction.
\sv{\qed}
\end{proof}
}

Lemma~\ref{lem-n0} is the last part of the proof that makes use of the $C_5$-freeness of~$G$. From 
now on, we will not need to use the fact that $G$ is $C_5$-free. In particular, we will allow 
ourselves reduction operations, such as the neighborhood collapse, which do not preserve 
$C_5$-freeness. 

We will assume, without mentioning explicitly, that after performing any modification of the 
list-3-coloring instance $G$, we always apply basic reductions until no more basic reductions are 
available.

In the rest of the proof, we use the term \emph{top component} to refer to a connected component of 
$N_2$. Observe that every top component is $P_4$-free and therefore has a dominating set of size at most 2~\cite{CP84}.
 We say that a top component is \emph{relevant}, if it contains a vertex $z$ with $|P(z)|=3$. 
Note that if $G$ has no relevant top component, then all its vertices have at most two available 
colors, and the coloring problem can be solved by a single basic reduction.
\lv{%

}%
We will say that a vertex $x$ of $N_1$ is \emph{relevant} if $x$ is adjacent to a vertex belonging 
to a relevant top component.
\lv{%

}%
Let $x\in N_1$ be a vertex, and let $C$ be a top component. We say that $x$ is a \emph{partial 
neighbor} of $C$, if $x$ is adjacent to at least one but not all the vertices of~$C$. We say that 
$x$ is a \emph{full neighbor} of $C$, if it is adjacent to every vertex of~$C$.

\begin{lemma}\label{lem-partial} \sv{($\clubsuit$)}
Suppose $x\in N_1$ is a partial neighbor of a top component $C$. Then $x$ is not a neighbor 
of any other top component. Moreover, $|N_0(x)|=2$.
\end{lemma}
\toappendix{
\lv{\begin{proof}}
  \sv{\begin{proof}[of Lemma~\ref{lem-partial}]}
Let $y$ and $z$ be two adjacent vertices belonging to $C$, such that $x$ is adjacent to $y$ but not 
to~$z$. Suppose for contradiction that there is a vertex $w\in N_2$ adjacent to $x$ but not 
belonging to~$C$. Then $wxyz$ is a copy of $P_4$ with three vertices in $N_{2}$, as shown in Figure~\ref{fig:N0x2}, which contradicts
Lemma~\ref{lem-n0}. This shows that $x$ is not adjacent to any top component other than~$C$.

Suppose now that $N_0(x)$ contains a single vertex $v_i$. Then $v_ixyz$ together with 
$v_{i+2}v_{i+3}v_{i+4}v_{i+5}$ induce a~$2P_4$.
\sv{\qed}
\end{proof}
}

We will now reduce $G$ to a set of polynomially many instances in which the set of relevant 
vertices has special form. We first eliminate the relevant vertices that have only one neighbor 
in~$N_0$.
Let $R_i$ be the set of relevant vertices that are adjacent to $v_i$ and not adjacent to any other 
vertex of~$N_0$. 

\toappendix{
\begin{figure}
\begin{subfigure}{0.45\textwidth}
\centering
\includegraphics[width=0.95\textwidth]{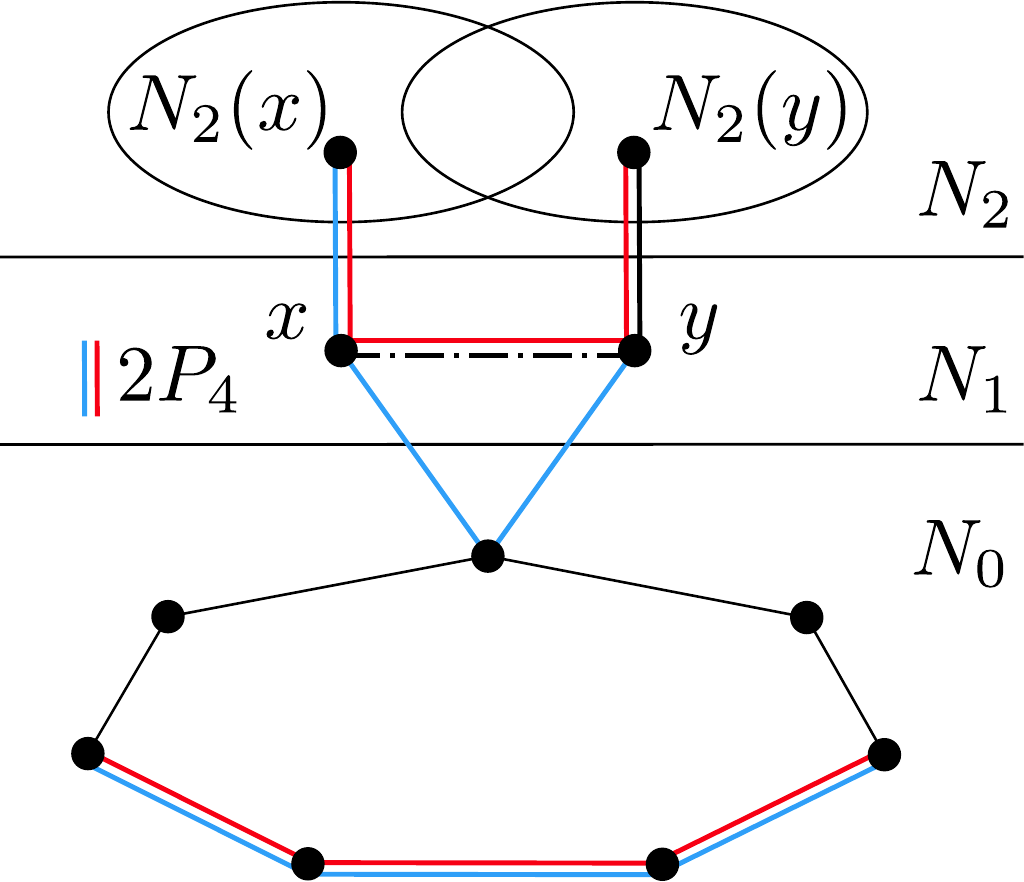}
\caption{The situation obtained from the assumption that $N_2(x)$ and $N_2(y)$ for $x,y \in R_i$ 
are not comparable by inclusion. The dash-and-dot line represents an edge which is present in one 
case (red induced $2P_4$) and absent in the other (blue induced $2P_4$).}
\label{fig:Ri-empty-xy}
\end{subfigure}
~
\begin{subfigure}{0.45\textwidth}
\centering
\includegraphics[width=0.95\textwidth]{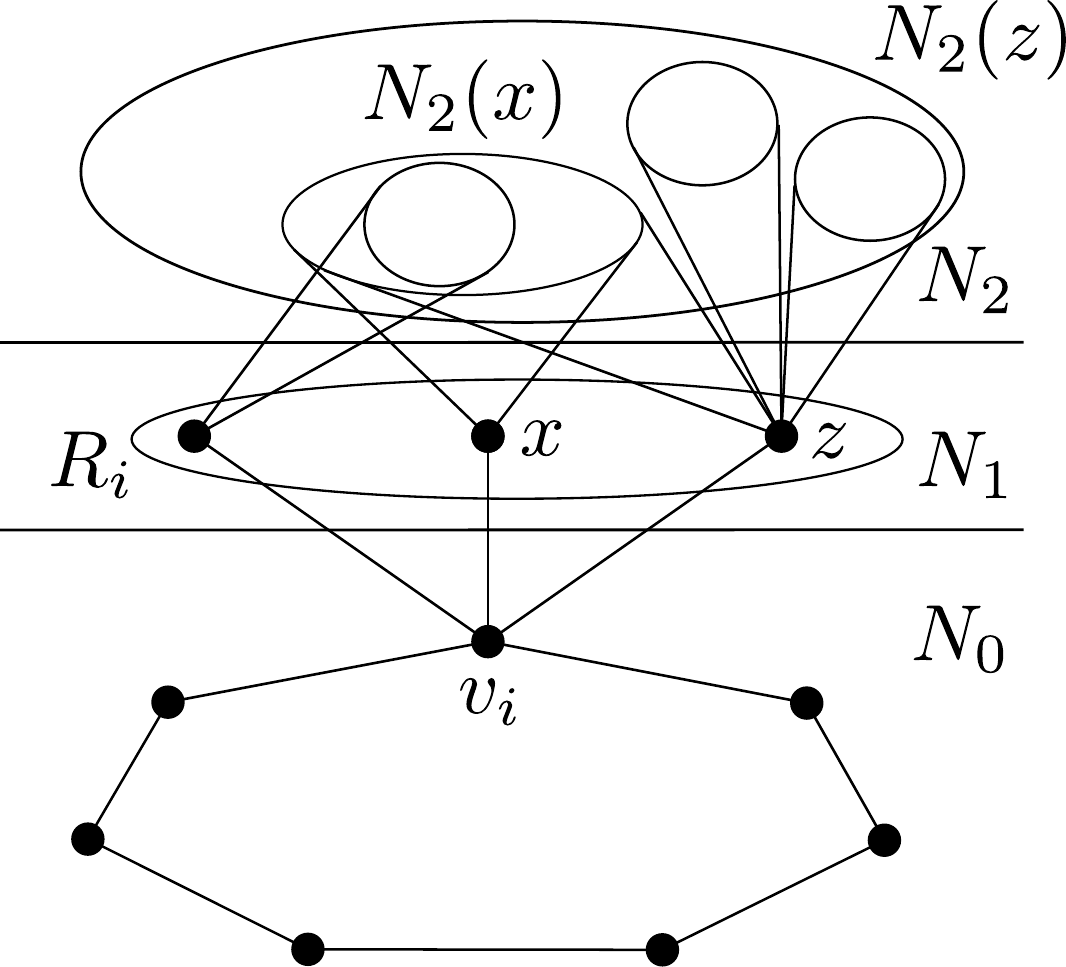}
\caption{The situation obtained when the neighborhoods of vertices in $R_i$ in $N_2$ are comparable by 
inclusion with $N_2(z)$ being the largest neighborhood. Note that these vertices are full neighbors 
of their top components.}
\label{fig:Ri-empty-xz}
\end{subfigure}
\caption{Illustrations of the situations in the proof of Lemma~\ref{lem-Ri}.}
\end{figure}
}

\begin{lemma}\label{lem-Ri}\sv{($\clubsuit$)}
 For any $i\in\{1,\dotsc,7\}$, we can reduce $G$ to an equivalent set of at most two instances, 
both of which satisfy $R_i=\emptyset$. 
\end{lemma}
\toappendix{
  \lv{\begin{proof}}
    \sv{\begin{proof}[of Lemma~\ref{lem-Ri}]}
By Lemma~\ref{lem-partial}, we know that any vertex $x\in R_i$ is a full neighbor of each of its
adjacent top components. 

Let $x,y$ be two distinct vertices of~$R_i$. We claim that the two sets $N_2(x)$ and $N_2(y)$ are 
comparable by inclusion. To see this, suppose for contradiction that there are vertices $x'\in 
N_2(x)\setminus N_2(y)$ and $y'\in N_2(y)\setminus N_2(x)$. Then we can find in $G$ a copy of 
$2P_4$ in which the first $P_4$ is $v_{i+2}v_{i+3}v_{i+4}v_{i+5}$, and the second $P_4$ is either 
$x'xyy'$ (if $xy\in E(G)$), or $x'xv_iy$ (if $xy\not\in E(G)$); see Figure~\ref{fig:Ri-empty-xy}.

Choose $z\in R_i$ so that $N_2(z)$ is as large as possible. In particular, for every $x\in R_i$, we 
have $N_2(x)\subseteq N_2(z)$. We then obtain two instances equivalent to $G$ by coloring $z$ by its 
two available colors. Note that by coloring $z$, we ensure that all the vertices $N_2(z)$ have at 
most two available colors, and since $z$ is a full neighbor of all its adjacent top components, 
this ensures that the vertices of $R_i$ will no longer be relevant after $z$ has been 
colored; see Figure~\ref{fig:Ri-empty-xz} for illustration.
\sv{\qed}
\end{proof}
}

From now on, we deal with instances of $G$ where every relevant vertex has exactly two neighbors 
in~$N_0$. Let $S_i$ be the set of relevant vertices adjacent to~$v_i$. 

\toappendix{
\bigskip
\begin{minipage}{0.55\textwidth}
\centering
\includegraphics[width=0.95\textwidth]{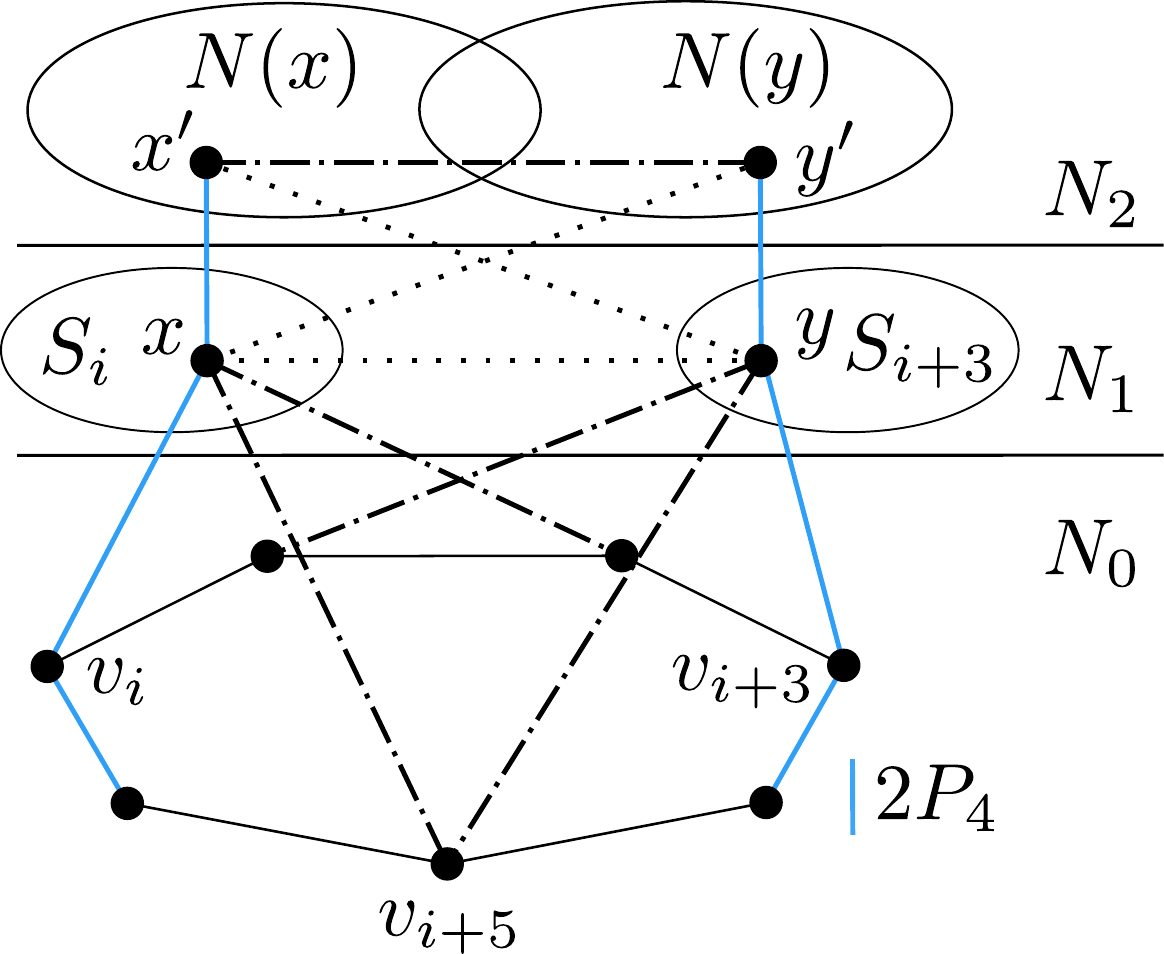}
\captionof{figure}{Considering a pair of vertices $(x,y)$ of type $\gamma$ for $x \in S_i, y \in 
S_{i+3}$, the edge $x'y'$ must be present, otherwise we obtain an induced $2P_4$. The other 
dash-and-dotted edges are not necessarily present, and the vertex $v_{i+5}$ is adjacent to at most 
one vertex from $\{x, y\}$.}
\label{fig:SiSi+3-xy}
\end{minipage}
~
\begin{minipage}{0.35\textwidth}
\centering
\includegraphics[width=0.95\textwidth]{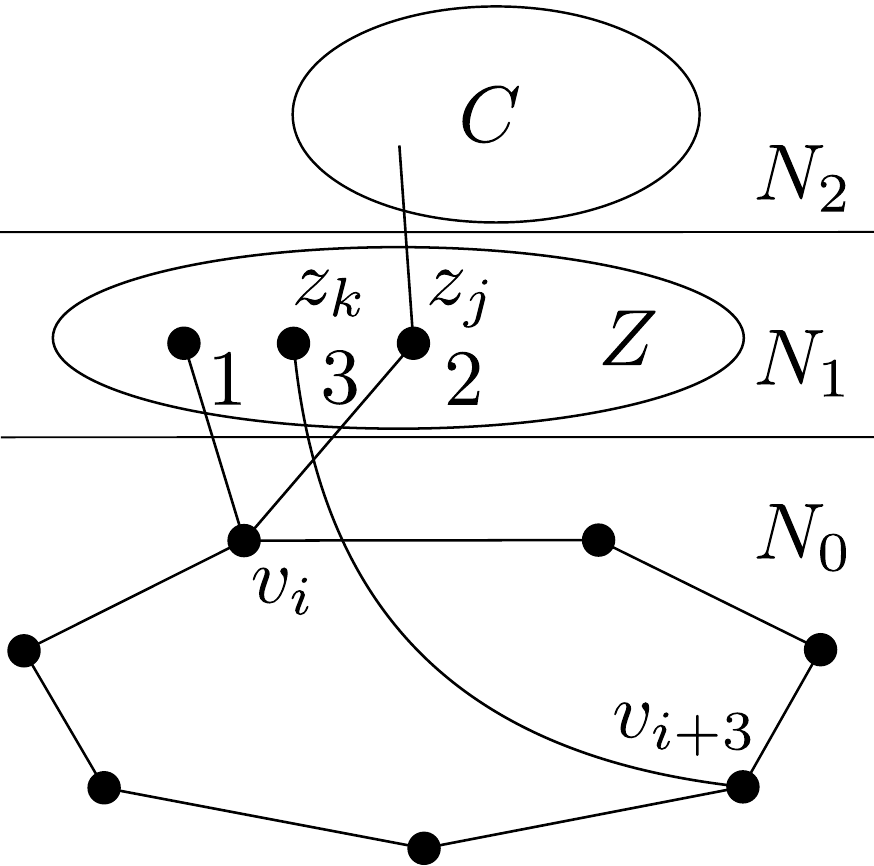}
\captionof{figure}{Coloring $(b)$ from the proof of Lemma~\ref{lem-SiSi+3empty} for $k<j$.}
\label{fig:SiSi+3-k<j}
\end{minipage}
\bigskip
}

\begin{lemma}\label{lem-SiSi+3empty} \sv{($\clubsuit$)}
For any $i\in\{1,\dotsc,7\}$, we can reduce $G$ to an equivalent set of polynomially many instances, 
each of which satisfies $S_i=\emptyset$ or $S_{i+3}=\emptyset$.
\end{lemma}

\toappendix{
  \lv{\begin{proof}}
    \sv{\begin{proof}[of Lemma~\ref{lem-SiSi+3empty}]}
Suppose that the vertices in $S_i$ have available colors 1 and 2, while the vertices in 
$S_{i+3}$ have available colors 2 and 3 (the case when the vertices in $S_{i+3}$ have the same 
available colors as the vertices in $S_i$ is similar and we omit it).

For a pair of vertices $x\in S_i$ and $y\in S_{i+3}$, we distinguish the following three 
possibilities:
\begin{itemize}
\item[($\alpha$)] $N_2(x)$ and $N_2(y)$ are comparable by inclusion,
\item[($\beta$)] $x$ is adjacent to $y$, or
\item[($\gamma$)] neither of the previous two conditions holds.
\end{itemize}
We say that the pair $(x,y)$ is of \emph{type $\alpha$} if it satisfies the condition $(\alpha)$ 
above, and similarly for the other two types. Observe that if the pair $(x,y)$ is of type~$\gamma$, 
then there exist $x'\in N_2(x)\setminus N_2(y)$ and $y'\in N_2(y)\setminus N_2(x)$. Moreover, for 
any choice of such $x'$ and $y'$, the pair $x'y'$ must be an edge of $G$, otherwise $x'xv_i 
v_{i-1}$ and $y'yv_{i+3}v_{i+4}$ would form a copy of~$2P_4$. In particular, $x'$ and $y'$ belong 
to the same top component $C$, and both $x$ and $y$ are partial neighbors of~$C$, as is depicted in Figure~\ref{fig:SiSi+3-xy}.

Let $Z=S_i\cup S_{i+3}$. Let $m$ be the size of $Z$, and let us order the vertices of $Z$
into a sequence $z_1,z_2,\dotsc,z_m$ satisfying $|N_2(z_1)|\ge |N_2(z_2)|\ge \dotsb\ge|N_2(z_m)|$.

We will reduce $G$ to the set of all the instances that can be constructed by one of the following 
two rules:
\begin{itemize}
\item[(a)] All the vertices in $Z$ are colored by their available color different from 2 (i.e., the 
vertices of $S_i$ are colored by 1, the vertices of $S_{i+3}$ by 3). 
\item[(b)] Fix a $j\in\{1,\dotsc,m\}$ and proceed as follows: color the vertices 
$z_1,\dotsc,z_{j-1}$ by their available color different from 2, and color $z_j$ by~2. Moreover, 
if $z_j$ is a partial neighbor of a top component $C$, color a dominating set of size two 
in~$C$, in all the possible ways. 
\end{itemize}
We now verify that in all the colorings described above, after all possible basic reductions are 
applied, either $S_i$ or $S_{i+3}$ becomes empty. This is clearly the case for the coloring described 
in~(a), in which all the vertices in $S_i\cup S_{i+3}$ will be removed from $G$, so both sets will be 
empty. 

Consider now a coloring described in~(b), and assume without loss of generality that $z_j$ is 
in~$S_i$. We claim that after the coloring is performed, there will be no relevant vertex left in 
$S_{i+3}$. To see this, consider a vertex $z_k\in S_{i+3}$. If $k<j$, then $z_k$ has been colored by 
the color~3, see Figure~\ref{fig:SiSi+3-k<j}. 

If $k>j$ we distinguish three possibilities depending on the type of the pair $(z_j,z_k)$. If the 
pair $(z_j,z_k)$ is of type $\alpha$, then $N_2(z_k)\subseteq N_2(z_j)$ (recall that $k>j$ implies 
$|N_2(z_j)|\ge |N_2(z_k)|$). In particular, all the vertices in any top component adjacent to $z_k$ 
will only have two available colors (recall that if $z_j$ is a partial neighbor of a top component, 
we also color a dominating set of this top component, ensuring all its vertices have at most two 
available colors). Thus, $z_k$ will no longer be relevant. If the pair $(z_j,z_k)$ is of type 
$\beta$, i.e. $z_j z_k$ is an edge, then $z_k$ has only the color 3 available and can be colored. 
Finally, suppose $(z_j,z_k)$ is of type~$\gamma$. As discussed before, this means both 
$z_j$ and $z_k$ are partial neighbors of a top component $C$ and have no other neighbors in~$N_2$. 
After the coloring is performed, all the vertices in $C$ will have only two available colors, 
because we have colored its dominating set of size two. Hence $z_k$ is no longer relevant. We 
conclude that $S_{i+3}$ becomes empty, as claimed.

It is clear that the coloring rules (a) and (b) admit only polynomially many possible colorings, and 
that any valid list coloring of $G$ extends one of the partial colorings described in (a) or in (b). 
Thus, we reduced $G$ to an equivalent set of polynomially many instances.
\sv{\qed}
\end{proof}
}

From now on, assume that we deal with an instance $G$ in which for every $i$, one of the two sets 
$S_i$ and $S_{i+3}$ is empty. Unless the instance is already completely solved, there must be 
at least one relevant vertex. Assume without loss of generality that $G$ has a relevant vertex
adjacent to $v_1$ and~$v_3$. It follows that $S_1$ and $S_3$ are nonempty, and hence $S_4$, $S_5$, 
$S_6$ and $S_7$ are empty. Moreover, as any relevant vertex is adjacent to a pair of vertices of the 
form $\{v_i,v_{i+2}\}$, it follows that $S_2$ is empty as well. In particular, every relevant 
vertex $x$ satisfies $N_0(x)=\{v_1,v_3\}$. It follows that all the relevant vertices have the same 
palette of size 2; assume without loss of generality that this palette is $\{1,2\}$.

We will now focus on describing the structure of the subgraph of $G$ induced by the relevant 
vertices and the relevant top components adjacent to them. Let $R$ denote the set of relevant 
vertices. Note that the subgraph of $G$ induced by $R\cup N_2$ does not contain $P_4$, 
otherwise we could use the path $v_4v_5v_6v_7$ to get a $2P_4$ in~$G$. 
\lv{%

}%
Note also that if two relevant vertices $x$ and $y$ are adjacent, then any common neighbor of $x$ 
and $y$ must be colored by color 3, thanks to the diamond consistency rule. We thus know that 
adjacent relevant vertices have no common neighbors outside~$N_0$. We may also assume that the graph 
induced by the relevant vertices is bipartite, otherwise $G$ would clearly not be 3-colorable.

\lv{
\begin{figure}
\centering
\includegraphics[width=0.5\textwidth]{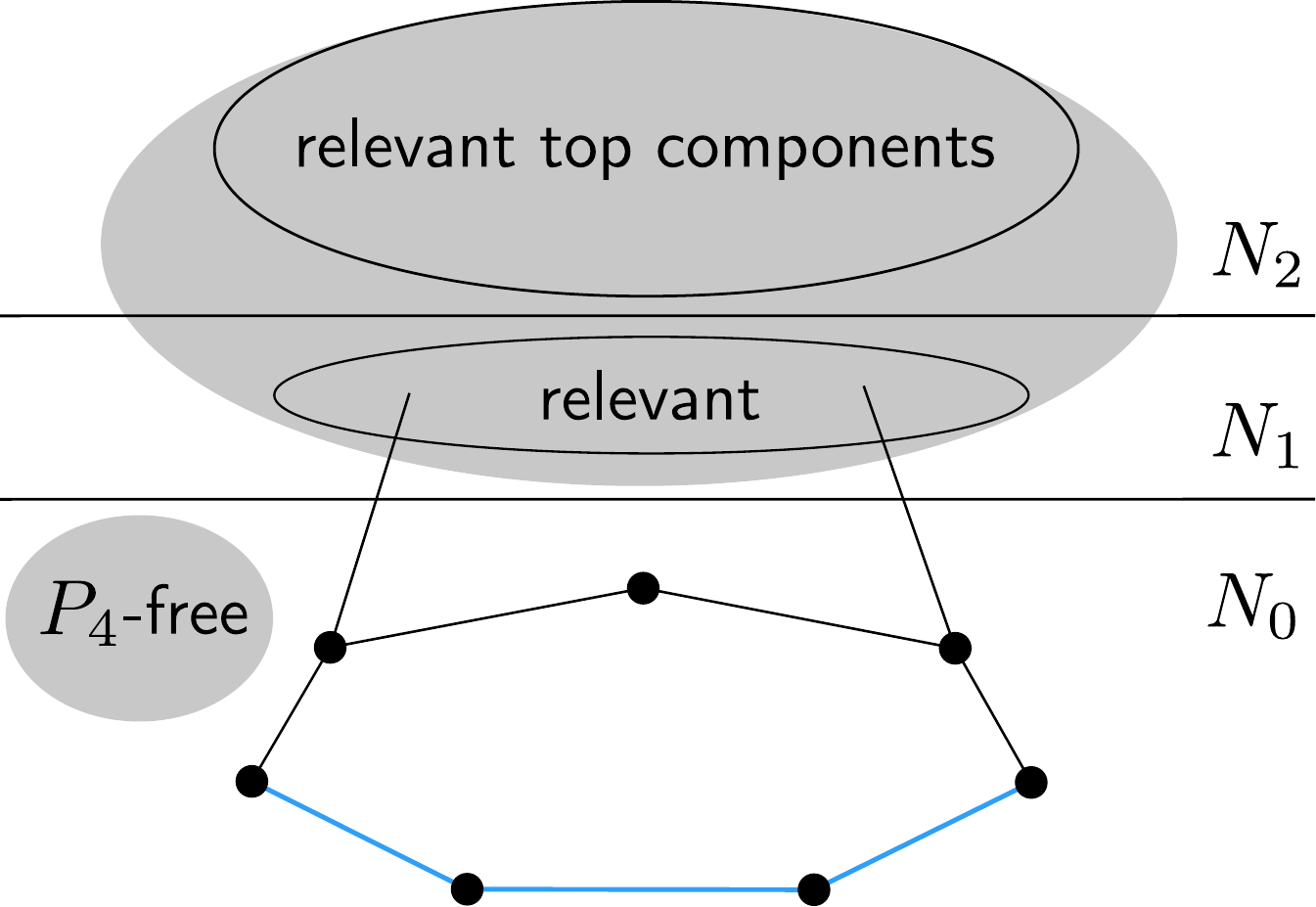}
\caption{The situation considered in the remaining part of the main proof.}
\label{fig:before-lem-rel}
\end{figure}
}

\begin{lemma}\label{lem-rel}\sv{($\clubsuit$)}
Suppose that $x$ and $y$ are two adjacent relevant vertices. Let us write $X'=N_2(x)$ and 
$Y'=N_2(y)$. Then there are disjoint sets $X, Y\subseteq R$, with $x\in X$ and $y\in Y$, satisfying 
these properties:
\begin{enumerate}
\item Every vertex in $X\cup Y'$ is adjacent to every vertex in $Y\cup X'$.
\item $X$ and $Y$ are independent sets of $G$.
\item The vertices in $X'\cup Y'$ are only adjacent to vertices in $X\cup Y\cup X'\cup Y'$; in 
particular, $X'\cup Y'$ induce a top component.
\end{enumerate}
\end{lemma}
\toappendix{
  \lv{\begin{proof}}
    \sv{\begin{proof}[of Lemma~\ref{lem-rel}]}
Consider the subgraph $G[R]$ of $G$ induced by the relevant vertices, and let $C$ be the connected 
component of $G[R]$ containing $x$ and~$y$. Recall that $C$ must be bipartite.
We let $X$ and $Y$ be its partite classes containing $x$ and $y$, respectively.
Note that $C$ is complete bipartite, otherwise it would contain a~$P_4$.

We will now show that all the vertices in $X$ have the same neighbors in~$N_2$. Indeed, if we could 
find a pair of vertices $x_1, x_2\in X$ and a vertex $x'\in N_2(x_1)$ not adjacent to $x_2$, then 
$x' x_1yx_2$ would induce a~$P_4$. It follows that for every $x_1\in X$ we have $N_2(x_1)=X'$, and 
similarly for every $y_1\in Y$ we have $N_2(y_1)=Y'$.

We saw that adjacent relevant vertices have no common neighbors, so $X'$ and $Y'$ are disjoint. Every 
vertex in $X'$ must be adjacent to every vertex in $Y'$, for if there were nonadjacent vertices 
$x'\in X'$ and $y'\in Y'$, then $x'xyy'$ would induce a~$P_4$. This proves the first claim of the 
lemma. 

To prove the second claim, observe that $X$ and $Y$ are independent by construction. 

To prove the third claim, proceed by contradiction and assume that a vertex $x'\in X'\cup Y'$ is 
adjacent to a vertex $z$ not belonging to $X\cup Y\cup X'\cup Y'$. We may assume that $x'$ belongs 
to~$X'$. Necessarily, $z$ belongs to $R\cup N_2$, and $zx'xy$ induces a forbidden~$P_4$.
\sv{\qed}
\end{proof}
}

Suppose $G[R]$ contains at least one edge $xy$, and let $X,Y,X',Y'$ be as in the previous lemma. 
Note that there are only two possible ways to color $G[X\cup Y]$ -- either $X$ is colored 1 and $Y$ 
is colored 2, or vice versa. We can check in polynomial time which of these two colorings can be 
extended to a valid coloring of $G[X\cup Y\cup X'\cup Y']$. If neither of the two colorings 
extends, we reject $G$, if only one of the two coloring extends, we color $X\cup Y$ accordingly, 
and if both colorings extend, we remove the vertices $X'\cup Y'$ from $G$, resulting in a smaller 
equivalent instance, in which $X\cup Y$ are no longer relevant. Repeating this for every component 
of $G[R]$ that contains at least one edge, we eventually reduce the problem to an instance in which 
the relevant vertices form an independent set.
\lv{%

}%
From now on, we assume $R$ is independent in~$G$. For a vertex $x\in R$, let $C_2(x)$ denote the set 
of top components that contain at least one neighbor of~$x$. 

\begin{lemma}\label{lem-equiv} \sv{($\clubsuit$)}
For any two relevant vertices $x$ and $y$, we either have $C_2(x)=C_2(y)$, or $C_2(x)$ and $C_2(y)$ 
are disjoint.
\end{lemma}
\toappendix{
  \lv{\begin{proof}}
  \sv{\begin{proof}[of Lemma~\ref{lem-equiv}]}
Suppose the lemma fails for some $x$ and $y$. We may then assume that there is a top component $C\in 
C_2(x)\cap C_2(y)$ and a component $C'\in C_2(x)\setminus C_2(y)$. Since $|C_2(x)|\ge 2$, we 
know from Lemma~\ref{lem-partial} that $x$ is a full neighbor of all the top components in~$C_2(x)$. 
Choose a vertex $u\in C'$ and a vertex $v\in C\cap N_2(y)$. Then $uxvy$ is a copy of $P_4$ in $R\cup 
N_2$, which is impossible.
\sv{\qed}
\end{proof}
}

Let us say that two relevant vertices $x$ and $y$ are \emph{equivalent} if $C_2(x)=C_2(y)$. As the 
next step in our algorithm, we will process the equivalence classes one by one, with the aim to 
reduce the instance $G$ to an equivalent instance in which each relevant vertex is adjacent to a 
single top component.

Let $x\in R$ be a vertex such that $|C_2(x)|\ge 2$, and let $R_x$ be the equivalence class 
containing~$x$. By Lemma~\ref{lem-partial}, each vertex in $R_x$ is a full neighbor of any component 
in $C_2(x)$, and by Lemma~\ref{lem-equiv}, no vertex outside of $R_x$ may be adjacent to a relevant 
top component in~$C_2(x)$. Thus, $R_x$ is a vertex cut separating the relevant top components in 
$C_2(x)$ from the rest of~$G$. We may therefore apply the cut reduction through the vertex cut $R_x$ 
to reduce $G$ to a smaller instance in which the vertices of $R_x$ are no longer relevant.

We repeat the cut reductions until there is no relevant vertex adjacent to more than one top 
component. From now on, we deal with instances in which each relevant vertex is adjacent to a unique 
top component; note that this top component is necessarily relevant. 

\toappendix{
\begin{figure}
\begin{subfigure}{0.45\textwidth}
\centering
\includegraphics[width=0.95\textwidth]{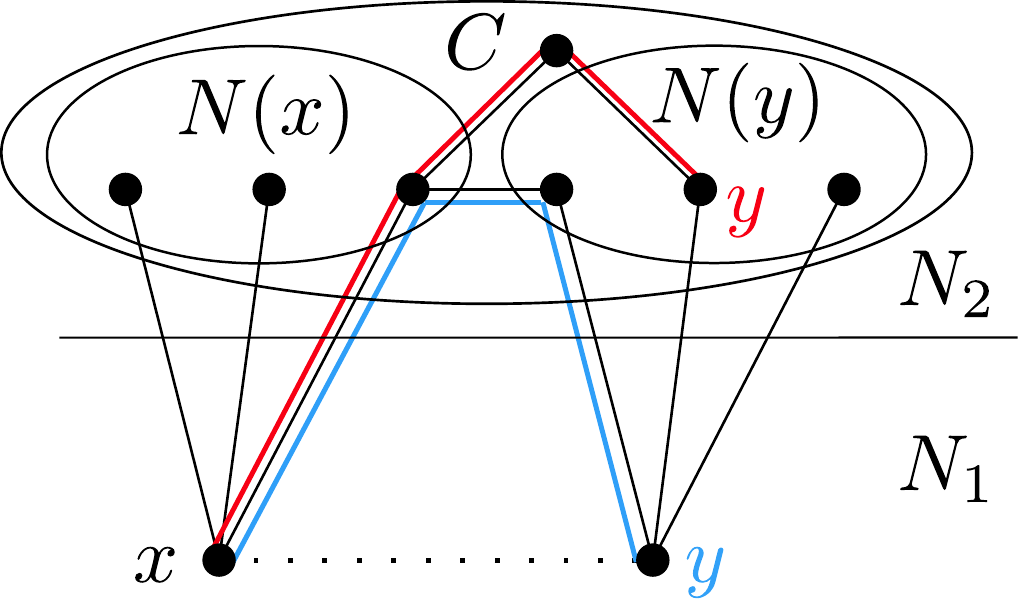}
\label{fig:lem-CN-1}
\end{subfigure}
~
\begin{subfigure}{0.45\textwidth}
\centering
\includegraphics[width=0.95\textwidth]{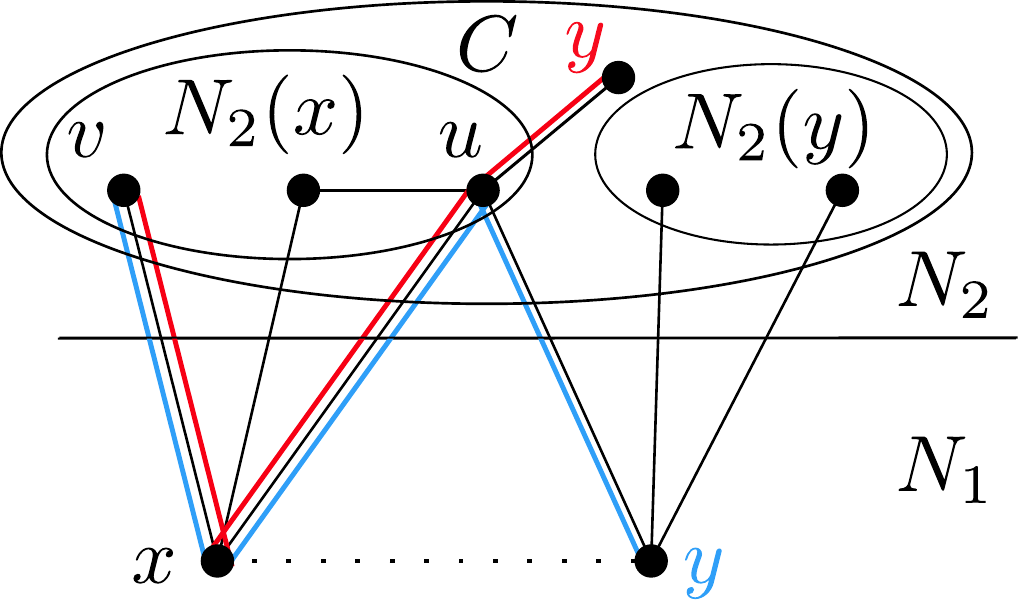}
\label{fig:lem-CN-2}
\end{subfigure}
\caption{Illustrations to the proof of Lemma~\ref{lem-CN}. The left part shows a situation when $y$ is not adjacent to any vertex in $N_2(x)$, the right part shows a situation when $y$ has a neighbour in $N_2(x)$ which is disconnected. 
Each part depicts two different possibilities.
The blue $P_4$ shows the case $y \in R_x$, while the red $P_4$ shows the case when $y\in C$.}
\label{fig:lem-CN}
\end{figure}
}

\begin{lemma}\label{lem-CN}\sv{($\clubsuit$)}
Let $x$ be a relevant vertex, let $C$ be the top component adjacent to $x$,
let $R_x$ be the equivalence class of $x$, and 
let $y\in R_x\cup C$ be a vertex not adjacent to~$x$. Then $y$ is adjacent to at least one vertex in 
$N_2(x)$. Moreover, if $N_2(x)$ induces a disconnected subgraph of $G$,
then $y$ is adjacent to all the vertices of~$N_2(x)$.  
\end{lemma}
\toappendix{
  \lv{\begin{proof}}
    \sv{\begin{proof}[of Lemma~\ref{lem-CN}]}
Refer to Figure~\ref{fig:lem-CN}.
If $y$ is not adjacent to any vertex of $N_2(x)$, then we can find an induced path with at least 
four vertices by considering the shortest path from $x$ to $y$ in the graph induced by 
$C\cup\{x,y\}$. Therefore $y$ has at least one neighbor in~$N_2(x)$. Suppose now that $N_2(x)$ is 
disconnected. If $y$ is not adjacent to all the vertices of $N_2(x)$, then we can find a vertex 
$u\in N_2(x)$ adjacent to~$y$, and a vertex $v\in N_2(x)$ nonadjacent to~$y$, in such a way that $u$ 
and $v$ are in distinct components of $N_2(x)$. Then $yuxv$ is an induced~$P_4$.
\sv{\qed}
\end{proof}
}

Fix now a relevant top component $C$ and let  
$R$ be set of relevant vertices in %
$N_1$ adjacent to~$C$. Fix a vertex $x\in R$ so that $N_2(x)$ is as large as possible.
Let $R_x$ be the equivalence class containing $x$.
We distinguish several possibilities, based on the structure of~$N_2(x)$.

\lv{\paragraph{$N_2(x)$ is disconnected.}}
\sv{\smallskip\noindent\textit{$N_2(x)$ is disconnected.~~}}
Suppose first that $N_2(x)$ induces in $G$ a disconnected subgraph. By Lemma~\ref{lem-CN}, any vertex 
in $R_x$ is adjacent to all vertices in $N_2(x)$. By our choice of $x$, this implies that for 
any $x'\in R_x$ we have $N_2(x')=N_2(x)$. We may therefore apply the cut reduction for the cut $R_x$ 
that separates $C$ from the rest of $G$, to obtain a smaller instance in which the vertices of $R_x$ 
are no longer relevant.

\lv{
\begin{minipage}{0.50\textwidth}
\centering
\includegraphics[width=0.95\textwidth]{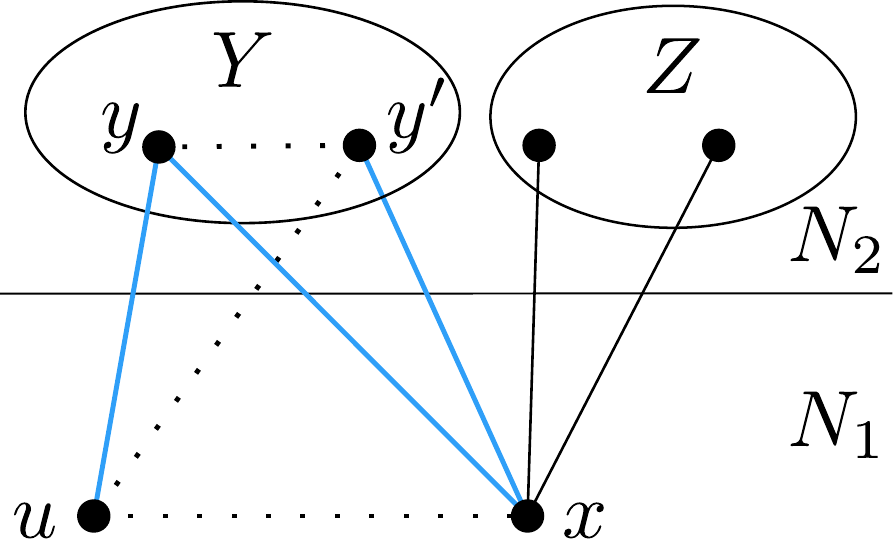}
\captionof{figure}{There is an induced $P_4$ in $N_2$ if $N_2(x)$ is connected with $\geq 3$ vertices 
and for $y, y' \in Y$ there exists a $u$ neighboring only one of them.}
\label{fig:N2-conn}
\end{minipage}
~
\begin{minipage}{0.40\textwidth}
\centering
\includegraphics[width=0.6\textwidth]{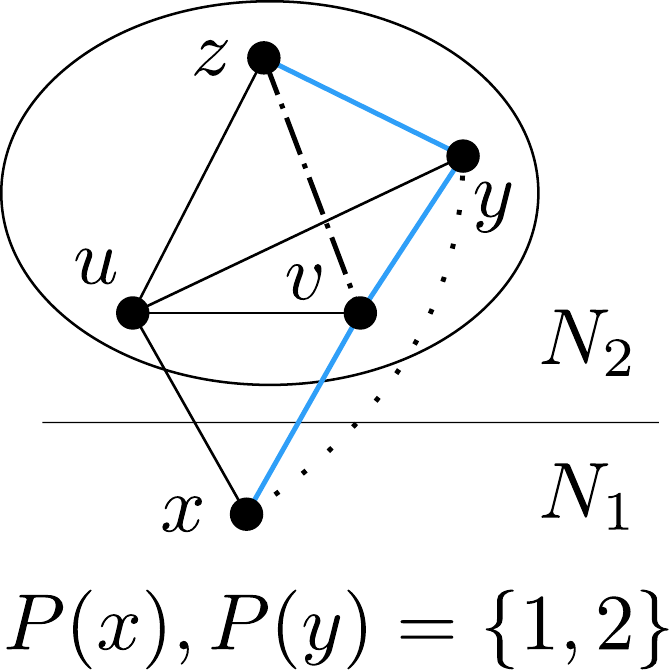}
\captionof{figure}{Recall that $u$ is adjacent to all vertices of $C$.
If there is a $y \in C$ adjacent to both $u$ and $v$, there is no other 
neighbor $z$ of~$y$. 
Otherwise, either $zv$ is an edge, causing 
a $K_4$, or $zv$ is not an edge, causing an induced $P_4$.}
\label{fig:uy-vy-edge}
\end{minipage}
}

\lv{\paragraph{$N_2(x)$ is connected, with $\ge3$ vertices.}}
\sv{\smallskip\noindent\textit{$N_2(x)$ is connected, with $\ge3$ vertices.~~}}
Now suppose that $N_2(x)$ induces a connected graph, and that $N_2(x)$ has at least three vertices. 
We now verify that $N_2(x)$ induces a complete bipartite graph, otherwise $C$ contains $P_4$ or $G$ is 
not 3-colorable. Let $Y$ and $Z$ be the two partite classes of $N_2(x)$. Note that any two vertices 
$y,y'$ in $Y$ have the same neighbors in $G$: indeed if $u$ were a vertex adjacent to $y$ but not 
to $y'$, then $uyxy'$ would induce a copy of~$P_4$\lv{, as depicted in Figure~\ref{fig:N2-conn}}. By the 
same argument, all the vertices 
in $Z$ have the same neighbors in $G$ as well. Diamond consistency enforces that all the vertices in 
$Y$ have the same palette, and similarly for~$Z$. We may then invoke neighborhood domination to 
delete from $Y$ all vertices except a single vertex $y$, and do the same with $Z$, reducing $G$ to an 
equivalent instance in which $N_2(x)$ consists of a single edge. 

\lv{\paragraph{$N_2(x)$ is a single vertex.}}
\sv{\smallskip\noindent\textit{$N_2(x)$ is a single vertex.~~}}
Suppose that $N_2(x)$ consists of a single vertex $y$. If $y$ is the only vertex of $C$, then $y$ 
must have the palette $\{1,2,3\}$, otherwise $C$ would not be a relevant component. In such case, we 
may simply color $y$ with color 3 and delete it, as this does not restrict the possible colorings of 
$G-y$ in any way. If, on the other hand, $C$ has more than one vertex, it follows from 
Lemma~\ref{lem-CN} that all the vertices of $R_x$ are adjacent to $y$, and by the choice of $x$, 
every vertex in $R_x$ is adjacent to $y$ as its only neighbor in~$C$. We may then apply cut reduction 
for the cut~$R_x$. In all cases, we obtain a smaller equivalent instance, in which the vertices in 
$R_x$ are no longer relevant.
\lv{
\begin{figure}
\begin{subfigure}{0.45\textwidth}
\centering
\includegraphics[width=0.95\textwidth]{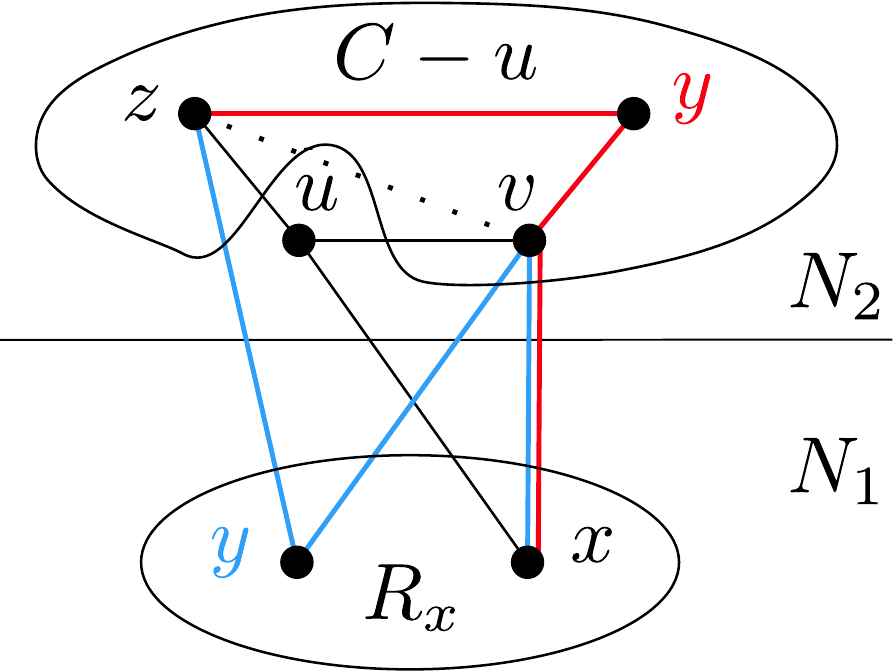}
\end{subfigure}
~
\begin{subfigure}{0.45\textwidth}
\centering
\includegraphics[width=0.95\textwidth]{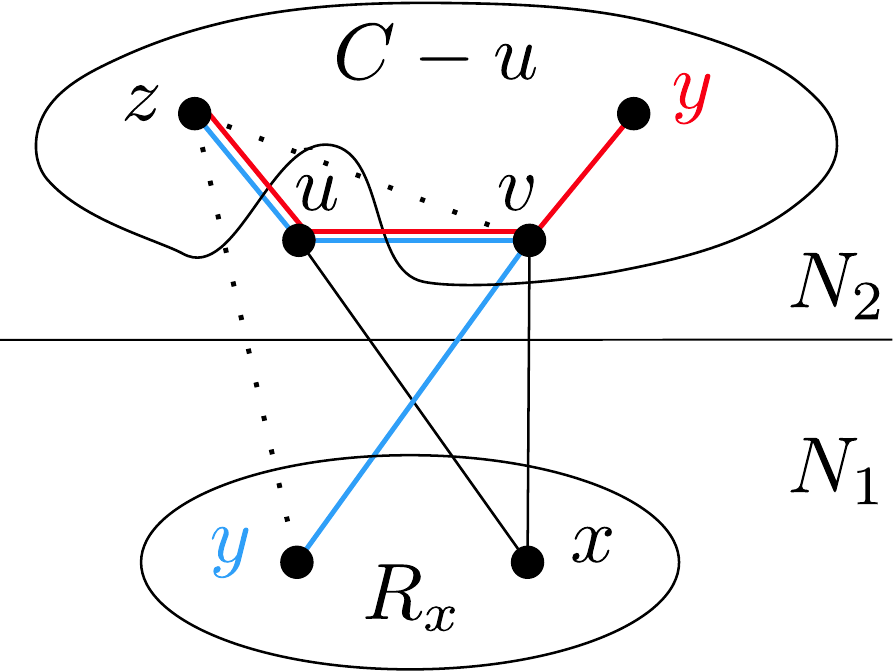}
\end{subfigure}
\caption{Illustrations of the situation when $N_2(x)$ is a single edge $uv$. Again, case $y \in R_x$ 
is shown as blue $y$ and blue $P_4$, while $y \in (C-u)$ is shown as red $y$ and red $P_4$. There 
are two subcases corresponding to $yz$ being an edge or not.}
\label{fig:N2-edge}
\end{figure}
}

\lv{\paragraph{$N_2(x)$ is a single edge.}}
\sv{\smallskip\noindent\textit{$N_2(x)$ is a single edge.~~}}
The last case to consider deals with the situation when $N_2(x)$ 
contains exactly two adjacent vertices $u$ and $v$. Assume that $\deg_G(u)\ge \deg_G(v)$. Recall 
that the set $R$ of relevant vertices is independent.
Note that for any vertex $x'\in R_x$, $N_2(x')$ is connected, otherwise Lemma~\ref{lem-CN} implies
that $N_2(x')$ is contained in $N_2(x)$, contradicting $N_2(x)$ being a single edge.

We first claim that any vertex $y\in R_x\cup (C-u)$ adjacent to $v$ is also adjacent to~$u$. Suppose 
this is not the case. Then, since $\deg_G(u)\ge \deg_G(v)$, there must also be a vertex $z\in R_x\cup 
(C-v)$ adjacent to $u$ but not to~$v$. If $yz$ is an edge, then $zyvx$ is a copy of $P_4$, and if 
$yz$ is not an edge, then $zuvy$ is a copy of~$P_4$\lv{, as shown in Figure~\ref{fig:N2-edge}}. In both cases we have a contradiction, 
establishing the claim. Note that the claim, together with Lemma~\ref{lem-CN}, implies that $u$ is 
adjacent to all the other vertices of $R_x\cup C$. 

Next, we show that if $C$ contains a vertex adjacent to both $u$ and $v$, then we may reduce $G$ to 
a smaller equivalent instance. Suppose $y\in C$ is adjacent to $u$ and~$v$. Then 
$P(y)=P(x)=\{1,2\}$ by diamond consistency. We now claim that $y$ has no other neighbors in 
$G$ beyond $u$ and~$v$. Suppose that $z\not\in\{u,v\}$ is a neighbor of~$y$. Then $z$ cannot be 
adjacent to $v$, since $uvyz$ would form a $K_4$.
Therefore $zyvx$ is a copy of $P_4$, a 
contradiction\lv{ illustrated by Figure~\ref{fig:uy-vy-edge}}. We conclude that $N(y)=\{u,v\}\subseteq N(x)$, and since $P(y)=P(x)$, we may delete 
$y$ due to neighborhood domination.

From now on, we assume that $u$ and $v$ have no common neighbor in~$C$. Recall that $u$ is adjacent 
to all the other vertices in $C\cup R_x$. We now reduce $G$ to an instance where $C-u$ is an 
independent set. We already know that $v$ is isolated in $C-u$ by the previous paragraph. Suppose 
that $C-u$ has a component $D$ with more than one vertex. If $D$ has a vertex $v'$ adjacent to a 
vertex $x'\in R_x$, we can repeat the reasoning of the previous paragraph with $x'$ and $v'$ in the 
place of $x$ and $v$, showing that $u$ and $v'$ cannot have any common neighbor in $C$, 
contradicting the assumption that $D$ has more than one vertex. We can thus conclude that $D$ is not 
adjacent to any 
vertex in~$R_x$. Then $u$ is a cut-vertex separating $D$ from the rest of~$G$. We may test which 
colorings of $u$ can be extended into $D$ (since $D$ is $P_4$-free, this can  be done efficiently), 
then restrict the palette of $u$ to only the feasible colors, and then delete~$D$.

\lv{
\bigskip
\begin{minipage}{0.35\textwidth}
\centering
\includegraphics[width=0.95\textwidth]{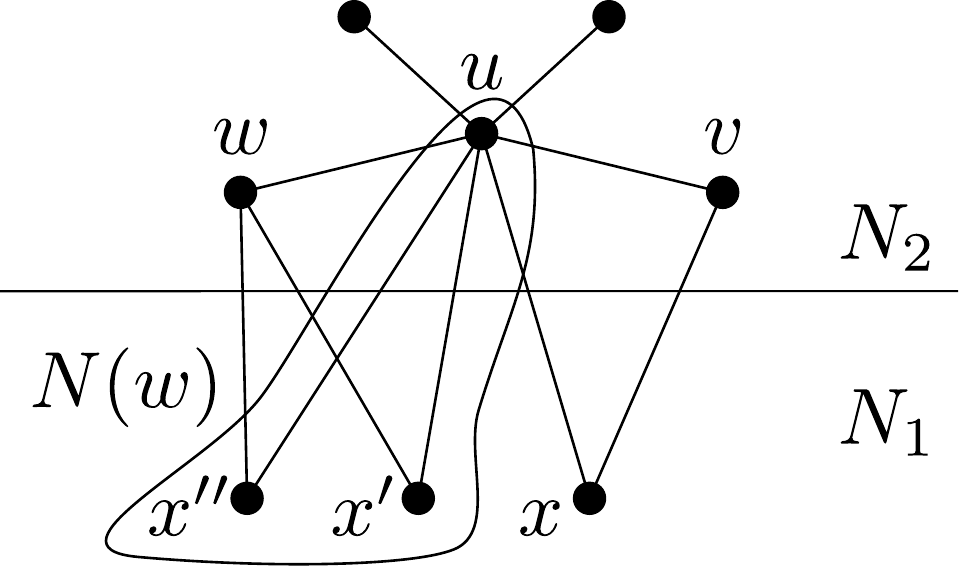}
\captionof{figure}{Situation in which we can apply a neighborhood collapse.}
\label{fig:star}
\end{minipage}
~
\begin{minipage}{0.55\textwidth}
\centering
\includegraphics[width=0.95\textwidth]{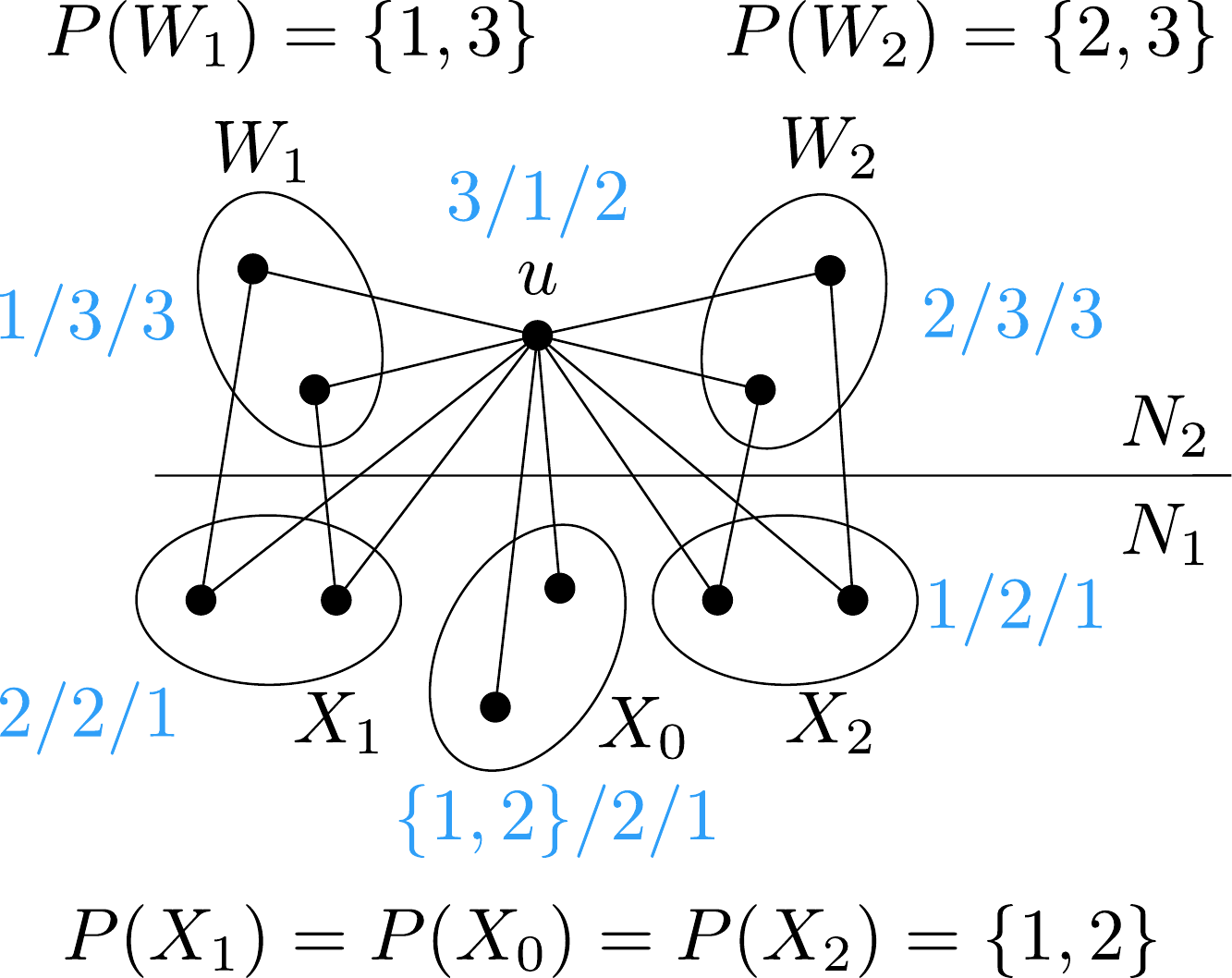}
\captionof{figure}{The last case in which each vertex of $C-u$ has palette $\{1,3\}$ or $\{2, 3\}$. The blue text represents the three possibilities to color $u$
and what colors that implies for other parts of the graph.}
\label{fig:WXu}
\end{minipage}
\bigskip
}

We are now left with a situation when $C$ is a star with center $u$, and every vertex of $R_x$ is 
adjacent to $u$ and to at most one vertex of~$C-u$. If there is a vertex $w\in C-u$ adjacent to more 
than one vertex in $R_x$, it means that the neighborhood of $w$ is a connected bipartite graph (a 
star with center $u$) to which we may apply neighborhood collapse\lv{; see Figure~\ref{fig:star}}.

Suppose now that every vertex $w\in C-u$ has only one neighbor in $R_x$ (if $w$ had no neighbor in 
$R_x$, it would have degree 1 and we could remove it). If $w$'s palette has 3 colors, we can remove 
it, so we may assume that every vertex in $C-u$ has a palette of size 2. Then $u$'s palette must 
have 3 colors, otherwise $C$ would not be a relevant component. If a vertex in $C-u$ has 
palette $\{1,2\}$, then $u$ must be colored 3 and then $R_x$ is no longer relevant. 

It remains to consider the case when each vertex of $C-u$ has the palette $\{1,3\}$ or~$\{2,3\}$. 
Let $W_1$ and $W_2$ be the sets of vertices of $C-u$ having palette $\{1,3\}$ and $\{2,3\}$, 
respectively. Let $X_1$ and $X_2$ be the sets of vertices of $R_x$ that are adjacent to a vertex in 
$W_1$ and $W_2$, respectively. Let $X_0$ be the set of vertices in $R_x$ that have no neighbor 
in~$C-u$.\lv{ The situation is shown in Figure~\ref{fig:WXu}.} Let us consider the possible colorings of 
$C\cup R_x$. If $u$ is colored by~3, then the 
whole set $W_1$ is colored by~1, $W_2$ is colored by~2, hence $X_1$ is colored by 2 and $X_2$ by 1, 
while the vertices in $X_0$ can be colored arbitrarily by 1 or~2. On the other hand, if $u$ receives 
a color $\alpha\neq 3$, then all the vertices in $R_x$ receive the color 
$\beta\in\{1,2\}\setminus\{\alpha\}$, and the vertices in $C-u$ can be colored by 3. The set $R_x$ 
therefore admits three types of feasible colorings: the all-1 coloring, the all-2 coloring, and any 
coloring where the set $X_1$ is colored by 2 and $X_2$ by~1. This set of colorings can be 
equivalently characterized by the following properties:
\begin{itemize}
\item If a vertex in $X_1$ is colored by 1, then the whole $R_x$ receives 1.
\item If a vertex in $X_2$ is colored by 2, then the whole $R_x$ is colored by 2.
\item All the colors in $X_1$ are equal and all the colors in $X_2$ are equal.
\end{itemize}
The above properties can be encoded by a 2-SAT formula whose variables correspond to vertices 
of~$R_x$.\sv{\qed}

\lv{
To summarize, we have shown that a 3-coloring instance $G$ can be reduced to an equivalent  
set of polynomially many simpler list-3-coloring instances. The structure of these simpler instances 
guarantees that for any relevant top component $C$, we can form a 2-SAT formula describing the 
colorings of the relevant vertices adjacent to $C$ that can be extended to a proper coloring of~$C$. 
Moreover, in the subgraph of $G$ induced by the vertices not belonging to any relevant top 
component, each vertex has a palette of size at most two. The colorings of this subgraph can again 
be encoded by a 2-SAT formula. Such an instance of list-3-coloring then admits a solution if and 
only if there is a satisfying assignment for the conjunction of the 2-SAT formulas described above. 
The existence of such an assignment can be found in polynomial time. This completes the proof of 
Theorem~\ref{thm:main}.
}

\lv{\section{Conclusions}}
\lv{
We have shown that 3-coloring on $(2P_4,C_5)$-free graphs is solvable in polynomial time.
As we discussed in the introduction, this approach might serve as a step towards resolving 
3-coloring on $2P4$-free graphs because it remains to consider $2P4$-free graphs containing $C_5$.

Apart from the main question above, under more refined scale, the complexity of 3-coloring on 
$(2P_4,C_3)$-free, $(P_8,C_3)$-free, or $(P_8,C_5)$-free graphs remains unknown.
In another direction, it would be interesting to extend our result to the list 3-coloring setting.
}

\lv{
\bigskip

\noindent\textbf{Acknowledgements.}~
We acknowledge the comfortable and inspiring atmosphere of the workshop KAMAK 2019 organized by 
Charles University where part of this work was done.
}

\lv{\bibliographystyle{plainurl}}
\sv{\bibliographystyle{splncs04}}
\bibliography{lit}

\end{document}